\documentclass[a4paper,fleqn,usenatbib]{mnras}

\usepackage{newtxtext,newtxmath}

\usepackage[T1]{fontenc}
\usepackage{ae,aecompl}

\usepackage{graphicx}	
\newsavebox\CBox

\usepackage{amsmath}	
\usepackage{amssymb}	

\usepackage{amsfonts}
\usepackage{bm}
\usepackage{hyperref}
\usepackage{threeparttable}

\title[CO(3--2) in hyper-luminous QSO fields]
{The SCUBA-2 Web Survey: I. Observations of CO(3--2) in hyper-luminous QSO fields}

\author[Hill et al.]{
\hspace{-5pt}Ryley Hill,$^{1,\thanks{ryleyhill@phas.ubc.ca}}$
Scott C.~Chapman,$^{1,2,3}$
Douglas Scott,$^{1}$
Ian Smail,$^{4}$
Charles C.~Steidel,$^{5}$\newauthor
Melanie Krips,$^{6}$
Arif Babul,$^{7}$
Trystyn Berg,$^{8}$
Frank Bertoldi,$^{9}$
Yu Gao,$^{10}$\newauthor
Kevin Lacaille,$^{11,3}$
Yuichi Matsuda,$^{12,13}$
Colin Ross,$^{3}$
Gwen Rudie,$^{14}$
Ryan Trainor$^{15}$
\\
$^{1}$Department of Physics and Astronomy, University of British Columbia, 6225 Agricultural Road, Vancouver, V6T 1Z1, Canada\\
$^{2}$National Research Council, Herzberg Astronomy and Astrophysics, 5071 West Saanich Road, Victoria, V9E 2E7, Canada\\
$^{3}$Department of Physics and Atmospheric Science, Dalhousie University, 6310 Coburg Road, Halifax, B3H 4R2, Canada\\
$^{4}$Centre for Extragalactic Astronomy, Department of Physics, Durham University, South Road,  Durham, DH1 3LE, UK\\
$^{5}$Cahill Center for Astronomy and Astrophysics, California Institute of Technology, MS 249-17, Pasadena, CA 91125, USA\\
$^{6}$Institut de Radio Astronomie Millim{\'e}trique, Domaine Universitaire, 300 Rue de la Piscine, Saint Martin d'H{\`e}res, F-38406, France\\
$^{7}$Department of Physics and Astronomy, University of Victoria, 3800 Finnerty Road, Victoria, V8P 1A1, Canada\\
$^{8}$European Southern Observatory, Alonso de C{\'o}rdova 3107, Santiago, Casilla 19001, Chile\\
$^{9}$Argelander-Institute f{\"u}r Astronomie, Rheinische Friedrich-Wilhelms Universit{\"a}t Bonn, Auf dem H{\"u}gel 71, Bonn, D-53121, Germany\\
$^{10}$Purple Mountain Observatory, Chinese Academy of Sciences, 8 Yuanhua, Nanjing, 2100034, China\\
$^{11}$Department of Physics and Astronomy, McMaster University, 1280 Main Street West, Hamilton, L8S 4M1, Canada\\
$^{12}$National Astronomical Observatory of Japan, Osawa 2-21-1, Mitaka, 181-8588, Japan\\
$^{13}$Graduate University for Advanced Studies (SOKENDAI), Osawa 2-21-1, Mitaka, 181-8588, Japan\\
$^{14}$Carnegie Observatories, 813 Santa Barbara Street, Pasadena, CA 91101, USA\\
$^{15}$Franklin \& Marshall College, Department of Physics and Astronomy, 415
Harrisburg Pike, Lancaster, PA 17603, USA
}

\date{8 February 2019}

\pubyear{2018}

\begin{document}
\label{firstpage}
\pagerange{\pageref{firstpage}--\pageref{lastpage}}
\maketitle

\begin{abstract}
\noindent
A primary goal of the SCUBA-2 Web survey is to perform tomography of the early inter-galactic medium by studying systems containing some of the brightest quasi-stellar objects (QSOs; 2.5$\,{<}\,z\,{<}\,$3.0) and nearby submillimetre galaxies. As a first step, this paper aims to characterize the galaxies that host the QSOs. To achieve this, a sample of 13 hyper-luminous ($L_{\rm AGN}\,{>}\,10^{14}\,{\rm L}_{\odot}$) QSOs with previous submillimetre continuum detections were followed up with CO(3--2) observations using the NOEMA interferometer. All but two of the QSOs are detected in CO(3--2); for one non-detection, our observations show a tentative 2$\sigma$ line at the expected position and redshift, and for the other non-detection we find only continuum flux density an order of magnitude brighter than the other sources. In three of the fields, a companion potentially suitable for tomography is detected in CO line emission within 25\,arcsec of the QSO. We derive gas masses, dynamical masses and far-infrared luminosities, and show that the QSOs in our sample have similar properties as compared to less luminous QSOs and SMGs in the literature, despite the fact that their black-hole masses (which are proportional to $L_{\rm AGN}$) are 1--2 orders of magnitude larger. We discuss two interpretations of these observations: this is due to selection effects, such as preferential face-on viewing angles and picking out objects in the tail ends of the scatter in host-galaxy mass and black-hole mass relationships; or the black hole masses have been overestimated because the accretion rates are super-Eddington.
\end{abstract}


\section{Introduction}


Star-forming galaxies (typically referred to as submillimetre galaxies, or SMGs) and quasi-stellar objects (QSOs, otherwise known as active galactic nuclei, or AGN) are responsible for maintaining the high ionization state of the inter-galactic medium (IGM) and for enriching the IGM with metals via strong supernovae (SNe), QSO-driven outflows, and `super-winds' (see \citealt{barkana2001} for a review of ionization of the IGM, and \citealt{veilleaux2005} for a review of galactic winds). Conversely, the physical conditions in the IGM (metallicity, over-density, etc.) are uniquely sensitive to the physics of the SNe and QSO feedback, which are believed to regulate the evolution of galaxies, particularly during the epoch when the density of star formation, mergers, and black hole accretion were at their highest. There is substantial empirical evidence for the impact of galaxies on the IGM, for example: the proximity effect of QSOs \citep[e.g.,][]{bajtlik1988,scott2000,aglio2008,perrotta2016}; powerful winds being driven off starburst galaxies \citep[e.g.,][]{fischer2010,spoon2013,cicone2014}; and radio jets \citep[e.g.,][]{bridle1984,zensus1997}.

There is also plenty of evidence supporting the claim that the supermassive black holes fuelling QSO activity grow in proportion to their host galaxies \citep[e.g.,][]{magorrian1998,gebhardt2000}, suggesting a tight link between SMGs and QSOs. This hypothesis, first proposed by \citet{sanders1988}, argues that a merger will initiate a large burst of star-formation, after which the gas can be funneled into the galaxy centre where it triggers QSO activity, and in some scenarios this will eventually quench the star-formation through a feedback cycle.

Testing this hypothesis requires observations at submillimetre (submm) wavelengths, as the copious amounts of dust generated from star-formation absorb the rest-frame optical and ultraviolet (UV) light emitted by stars and re-radiate it in the far-infrared (IR), observed here on Earth in the submm due to the high redshifts (usually $>2$) of these objects. For instance, one can compare the far-IR luminosities of SMGs and QSOs as a proxy of their total dust contents \citep[e.g.,][]{elbaz2010,hatziminaoglou2010,dai2012,leipski2014,verdier2016}. Observations of molecular gas lines (such as CO) in QSOs are also invaluable as they are associated with their host galaxies, allowing one to compare QSOs with their progenitor SMGs \citep{solomon2005,carilli2007,coppin2008,wang2010,simpson2012}; indeed, it is found that the molecular gas properties of local infrared QSOs, such as the molecular gas mass, star formation efficiency, and molecular linewidths are indistinguishable from those of ultraluminous infrared galaxies \citep[e.g.][]{xia2012}. In addition, one can look at the submm output of dust-obscured QSOs to search for evidence of recent mergers \citep[e.g.,][]{carilli2002,krips2005,riechers2008,salome2012,cariani2013}, and outflows can be observed in bright CO lines \citep[e.g.,][]{veilleux2013,sun2014,feruglio2015,spilker2018}.

The Keck Baryonic Structure Survey \citep[KBSS; e.g.,][]{rudie2012,trainor2012} is a unique spectroscopic survey designed to explore details of the connection between galaxies, QSOs, and their surroundings, and focuses particularly on scales from around 50\,kpc to a few Mpc. It includes a large sample of rest-frame-UV ($>5000$) and rest-frame-optical ($>1000$) spectra of UV-colour-selected star-forming galaxies at $z\approx2.3$. These galaxies were photometrically selected to lie in the foreground of 15 hyper-luminous ($L_{\rm AGN} > 10^{14}\,{\rm L}_{\odot}$) QSOs from 15 fields in the redshift range $2.5 < z < 3.0$, for which high-resolution, high signal-to-noise ratio (S/N) spectra had already been obtained, and were followed up extensively at various other wavelengths.

As part of the SCUBA-2 Web survey (PI: S.~Chapman), we have undertaken deep submm observations of the 15 KBSS fields at 850 and 450\,$\mu$m (Ross et al.~in preparation) using the James Clerk Maxwell Telescope (JCMT) Submillimetre Common User Bolometer Array-2 \citep[SCUBA-2;][]{holland2013}. The primary goals of the SCUBA-2 Web survey are: (1) to probe the far-IR luminosities and environments of the rare ultra-violet, hyper-luminous QSOs themselves; (2) to perform tomography of the IGM, probing the cosmic web using SMGs lying close to the luminous QSOs; and (3) to probe the bias of SMGs to UV/optical-selected galaxies, over a range of environments, obtained through comparison of the SMG redshifts to the extremely well sampled spectroscopic databases in these fields.

As a follow-up to this programme, we have targeted the molecular gas emission from 13 KBSS QSOs with the NOrthern Extended Millimeter Array \citep[NOEMA, developed from the Plateau de Bure Interferometer; see][]{guilloteau1992,chenu2016} in order to further characterize this sample of hyper-luminous QSOs using submm spectroscopy, which are amongst the brightest in the observable Universe. In particular, we would like to confirm the origin of the submm continuum emission seen by SCUBA-2 with higher resolution imaging, and to derive a variety of useful physical properties that can be compared to other populations of galaxies and QSOs; the NOEMA observations reported here will certainly reveal whether the molecular gas properties of these hyper-luminous QSOs are similar to that of the general population of SMGs. As another goal of the SCUBA-2 Web survey is to perform tomography of the IGM, we would also like to search for companion galaxies with suitable QSO allignments. These observations will help us to better understand the environments of these incredibly rare objects, including the mechanisms responsible for generating such vigorous QSOs and how they might affect their surroundings on scales much bigger than the sizes of their host galaxies.

In Section \ref{observations} of this paper we describe in detail the set of observations we have carried out using the NOEMA interferometer, the data reduction, and the methods we used to detect and characterize the spectra; in Section \ref{results} we derive various physical properties of our sample and compare the results with similar populations of objects found in the literature; in Section \ref{discussion} we discuss our data and provide plausible interpretations of our findings; and in Section \ref{conclusion} we summarize and conclude the paper. A Hubble constant $H_0=67.3$\ km\,s$^{-1}$\,Mpc$^{-1}$ and density parameters $\Omega_{\Lambda}=0.685$ and $\Omega_{\rm m}=0.315$ from \citet{planck2014-a15} are assumed throughout. For reference, at $z=2.7$, 1\,arcsec corresponds to about 8 physical kiloparsecs.

\section{Observing CO(3--2) in the KBSS QSOs}\label{observations}

\subsection{Sample selection}

The KBSS QSOs were originally selected as the brightest possible QSOs within the 3100--6000\,\AA\ range satisfying the following criteria: the redshifts needed to be between 2.55 and 2.85, important as this corresponds to the peak of star formation in the Universe \citep[e.g.,][]{sobral2013,koprowski2017}, shifts rest-frame UV light into the optical, and avoids contamination from the Lymann-$\alpha$ forest \citep[for more details see][]{steidel2014}; the fields needed to be low in foreground Galactic extinction; and the declinations needed to be greater than $-$20\,deg so as to be observable with Keck on Mauna Kea. This selection criteria resulted in a sample of 15 QSOs that were amongst the most UV-luminous quasars known.

Our ultimate goal was to observe CO\,$(J\,{=}\,3\rightarrow2)$ transition lines (rest-frame frequency 345.796\,Hz) in all 15 of the KBSS QSOs using the NOEMA interferometer. One source, HS1549$+$1919, had already been observed (in the summer 2016 semester) in both CO(3--2) and CO(7--6) (Chapman et al.~in preparation), revealing a nearby companion and strong evidence for a merger or massive disc through the existence of remarkably broad, 1500\,km\,s$^{-1}$ emission line; for comlpeteness, we have included HS1549$+$1919 in this study. One other source, Q0142$-$100, is a doubly-imaged gravitational lens separated by 2.2\,arcsec, with the brighter and fainter images being magnified by factors of approximately 3.2 and 0.4, respectively \citep{sluse2012}. This means that it is not as intrinsically luminous as it appears, but we have nevertheless kept it in the sample. We obtained time for the summer 2017 semester to observe the remaining 14 sources, but due to time and weather constraints, only 12 of these 14 fields were observed. We note that none of our targets are radio loud (meaning that the flux ratio of 5\,GHz to the $B$ band is greater than 10) in the Very Large Array (VLA) Faint Images of the Radio Sky at Twenty-cm (FIRST) survey \citep{gregg1996}, but all were known to be submm luminous based on SCUBA-2 imaging (Ross et al.~in preparation) at the time of the NOEMA observations. We emphasize that this had no impact on our NOEMA programme (e.g.~the two sources that were missed are similar to the others), and does not introduce additional selection effects into our sample.

\subsection{NOEMA observations}

Observations were taken using the NOEMA interferometer between June 2017 and August 2017 in Band 1 in the D (compact) configuration. In this configuration and at our observation frequency of about 100\,GHz, the primary beam has a full widths at half maximum (FWHM) of 50\,arcsec while the synthesized beam has a typical resolution of 3.7\,arcsec (this angular scale corresponds to about 30\,kpc at $z=2.7$). We tuned the observations to centre the CO(3--2) transition of each source in the middle of the lower sideband, typically around 95\,GHz. We used the WideX correlator with a standard spectral resolution of about 2\,MHz and an effective bandwidth of 3.6\,GHz. Depending on the local sidereal time (LST) range, one of the quasars 3C84, 3C273 or 2013$+$370 were used as bandpass calibrators. We observed one of the two strong radio stars MWC349 or LkH$\alpha$101 for absolute flux calibration; these two secondary flux calibrators are regularly monitored against planets and provide a flux density calibration accuracy of about 5--10 per cent in the lower 3-mm band. Phases and amplitudes were calibrated against quasars that were chosen within 10 degrees of the respective target and observed every 22 minutes, in alternation with the target. Depending on the target and weather conditions, we integrated between 2 and 5 hours on each source, resulting in a noise level in the range of 0.15--0.3\,mJy\,beam$^{-1}$ at a spectral resolution of 20\,MHz (corresponding to about 50 km\,s$^{-1}$ at the observed frequencies).

\subsection{Data reduction and source detection}\label{reduction}

\begin{figure*}
\centering
\makebox[\textwidth][c]{
\includegraphics[width=1.1\textwidth]{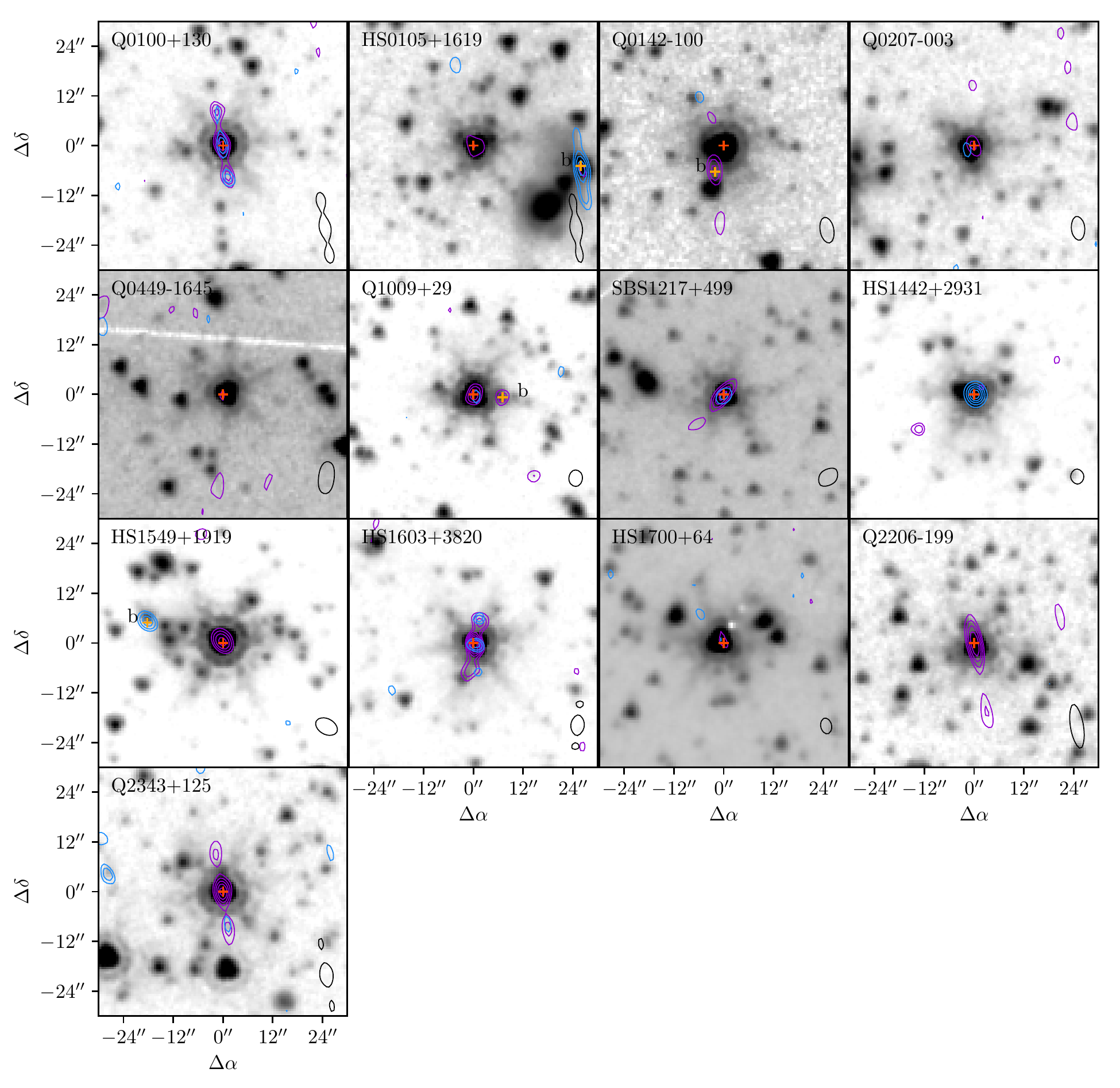}}
\caption{Dirty-map contours from our NOEMA observations plotted over IRAC 3.6-$\mu$m imaging around each KBSS QSO. The images are 60\,arcsec on a side; for reference, the NOEMA primary beam is 50\,arcsec at our observation frequency of about 100\,GHz. The purple contours are the mean of the frequency bins where a CO(3--2) line is seen, starting at 3$\sigma_{\rm CO}$ and increasing in steps of 1$\sigma_{\rm CO}$ (except for Q2343$+$123, which increases in steps of 2$\sigma_{\rm CO}$), where $\sigma_{\rm CO}$ is the standard deviation of the stacked CO(3--2) image. The blue contours show continuum emission and are the means of the frequency bins lacking line emission, starting from 3$\sigma_{\rm cont}$ and increasing in steps of 1$\sigma_{\rm cont}$ (except for HS1442$+$2931, which increases in steps of 2$\sigma_{\rm cont}$), where $\sigma_{\rm cont}$ is the standard deviation of the stacked continuum image. Red pluses indicate the positions of the peak CO(3--2) emission centroids (or, for sources lacking a line, continuum emission centroids), and orange pluses indicate the same for the companions. The dirty beams are shown in the bottom-right of each panel; the prominent sidelobes seen in Q0100$+$130, HS0105$+$1619, and HS1603$+$3820 are due to poor $uv$ sampling in some of our NOEMA observations.}
\label{fig:maps}
\end{figure*}

The NOEMA $uv$ tables were calibrated and imaged using the standard {\tt GILDAS} software. We set each map size to 128$\,{\times}\,$128 pixels with pixel sizes of 0.7$\,{\times}\,$0.7\,arcsec, and produced maps in units of Jy\,beam$^{-1}$. We chose to perform our analysis on dirty maps as opposed to {\tt CLEAN}ed maps, since dirty maps are known to be more reliable for interferometric data lacking extensive $uv$ coverage, due to contamination from imaging artefacts. Flux density uncertainties were estimated separately for each frequency bin by calculating the standard deviation of the dirty maps. We searched the resulting data cubes in the vicinity of the QSOs for clear line features at the expected frequencies, extracted 1D spectra from the highest S/N pixels in the dirty maps, then binned these by a factor of 2. Assuming our sources to be unresolved at the NOEMA spatial resolution, the units of the single-pixel spectra are then in Jy. 

In 12 of the 13 targets (including the existing data from HS1549$+$1919) we found line features at the location of the QSO and at the expected frequency. In order to assess the significance of these lines, we produced average (i.e.~moment 0) CO(3--2) maps by summing the frequency slices across the best-fit FWHM of the detected lines (see Section \ref{lines} for details on the fitting procedure); noise estimates were obtained by calculating their standard deviations. 10 of the 12 line features were found to be $>4\sigma$, one (Q0449$-$1645) was detected at $3\sigma$, and one (Q0142$-$100) at $2\sigma$. Since the positions and redshifts of all the QSOs in our sample are known {\it a priori}, we do not require very high significance in our lines and choose a detection threshold of 3$\sigma$, and we consider the 2$\sigma$ line of Q0142$-$100 to be tentative. Similarly, we produced continuum maps by summing the remaining frequency slices. In Fig.~\ref{fig:maps} we show the resulting CO(3--2) and continuum maps plotted over existing {\it Spitzer\/}-IRAC 3.6-$\mu$m maps, and in Fig.~\ref{fig:spectra_qso} we show the QSO spectra around the expected line velocities.

The source lacking a line, HS1442$+$2931, shows very bright continuum emission of about 1.6\,mJy across the band. For this QSO, we fit a phase-centred flat spectrum (i.e.~a polynomial of order zero) point source directly to the $uv$ data after masking the frequency range of the expected CO(3--2) line, and subtracted the best-fitting $uv$ points from the original data. However, the continuum-subtracted maps for HS1442$-$2931 did not contain any line features around the vicinity of the QSO. For the gravitational lens Q0142$-$100, which has two images, comparison with existing {\it Hubble\/} near-IR imaging shows that we have tentatively detected a CO(3--2) line only from the brighter of the two images, as the secondary image (located 2.2\,arcsec away from the primary) is outside of the NOEMA synthesized beam for this particular observation; however, we note that the astrometric precision of NOEMA is about 1\,arcsec, thus it is possible that the flux from both images has been combined in our data. The brighter image is known to be magnified by a factor of 3.2, while the fainter is magnified by a factor of $-$0.4, but we note that emission lines might be magnified by larger factors based on the size of the emitting region. For the purposes of this paper we will assume that we have only detected the brighter image and use a magnification factor of 3.2 for all further calculations, while we provide additional details of this system in Appendix \ref{appendix}.

From the continuum maps we measured continuum flux densities at the positions of the QSOs, and subtracted these from the spectra. We note that in several cases the continuum emission was consistent with 0, so only 3$\sigma$ upper limits could be derived for the actual source flux densities. In Table \ref{table:continuum} we provide these continuum flux density measurements, along with existing submm flux density measurements at 450 and 850\,$\mu$m, obtained from SCUBA-2 imaging (Ross et al.~in preparation).

We also searched our continuum and CO(3--2) maps for companion galaxies. As we were searching these maps well outside of the phase centres, we weighted the noise by the value of the normalized primary beam. In order to choose a S/N threshold we checked the negative peaks in all of our maps, and found none to be greater than a S/N of 4 within 60\,arcsec. By requiring a S/N$\,{>}\,$4 in either our CO(3--2) maps or in our continuum maps we recovered a total of three sources, two in a CO(3--2) map (Q0142$-$100 and Q1009$+$29) and one in a continuum map (HS1549$+$1919). In the HS0105$+$1619 field we have also detected in the continuum a known radio source from the NRAO VLA Sky Survey \citep[NVSS;][]{condon1998} radio continuum source catalogue at a S/N of 3.3, which we include here for completeness, but for the remainder of this paper we do not count it as a companion galaxy.

Without having checked other velocity slices within the bandwidth of our observations, the two sources detected at the same redshift as their field QSOs are somewhat questionable. For reference, the correlator bandwidth of 3.6\,GHz corresponds to a redshift range of about 0.1. We therefore tested their validity by randomly choosing 350\,km\,s$^{-1}$ velocity slices within the remaining bandwidth and repeating the search criteria outlined above. This was repeated 10 times for all 13 fields. In the end we found no additional negative or positive peaks with a S/N$\,{>}\,$4, making it much more likely that these sources are in fact galaxies associated with their field QSOs.

These companion sources are shown in Fig.~\ref{fig:maps} with a suffix `b', and in Fig.~\ref{fig:spectra_companion} we show the spectra of these sources in the vicinity of their field QSO redshifts. Looking at Fig.~\ref{fig:maps} we can see that Q1009$+$29b and HS1549$+$1919b have likely IRAC 3.6-$\mu$m counterparts, further improving their credibility; Q0142$-$100b is found between its corresponding QSO and a bright IRAC source, but it is clearly not associated to either; and due to saturation from the very bright and nearby QSO, it is impossible to identify any fainter IRAC sources at its position.

\begin{table}
\centering
\caption{Table of QSO submm continuum flux densities, with companions in italics and given a suffix `b'. Here the 3-mm continuum flux densities are more precisely at the frequencies $\nu_{\rm CO(3-2)}/(1+z)$, where $\nu_{\rm CO(3-2)} =$ 345.796\,GHz and the redshifts are provided in Table \ref{table:qso}. We also give 450\,$\mu$m and 850\,$\mu$m flux desnities from Ross et al.~(in preparation) using SCUBA-2. Where the continuum flux densities are consistent with 0, we provide 3$\sigma$ upper limits.}
\label{table:continuum}
\begin{threeparttable}
\begin{tabular}{|lccc|}
\hline
Name  & $S_{\rm 450\,\mu m}$ & $S_{\rm 850\,\mu m}$ & $S_{\rm 3\,mm}$ \\
 & [mJy] & [mJy] & [mJy] \\
 \hline
Q0100$+$130 & $<$35 & 6.0$\pm$1.2 &  0.22$\pm$0.04  \\
HS0105$+$1619 & $<$24 & 3.8$\pm$0.6 & 0.12$\pm$0.08 \\
\textit{HS0105$+$1619b} & $<$24 & $<$3.0 & 0.52$\pm$0.16 \\
Q0142$-$100 & $<$12$^{\rm a}$ & 2.2$\pm$0.4$^{\rm a}$ & $<0.03^{\rm a}$ \\ 
\textit{Q0142$-$100b} & $<$39 & $<$3.6 & 0.05$\pm$0.03 \\ 
Q0207$-$003 & 28.5$\pm$7.7 & 5.5$\pm$0.8 & 0.14$\pm$0.04 \\
Q0449$-$1645 & $<$46 & 4.7$\pm$0.9 & 0.05$\pm$0.05 \\
Q1009$+$29 & $<$38 & 8.9$\pm$1.1 & 0.12$\pm$0.04 \\
\textit{Q1009$+$29b} & $<$38 & $<$3.3 & 0.07$\pm$0.04 \\
SBS1217$+$499\phantom{0a} & 45.1$\pm$11.0 & 4.9$\pm$1.1 & 0.18$\pm$0.04 \\
HS1442$+$2931 & $<$42 & 2.8$\pm$1.0 & 1.64$\pm$0.12 \\ 
HS1549$+$1919 & 33.6$\pm$4.6 & 8.0$\pm$1.2 & 0.11$\pm$0.04 \\
\textit{Q1549$+$1919b} & $<$14 & $<$3.6 & 0.16$\pm$0.05 \\
HS1603$+$3820\phantom{a} & 33.4$\pm$10.8 & 7.7$\pm$1.2 & 0.15$\pm$0.03 \\
HS1700$+$64\phantom{00a} & 15.3$\pm$4.1 & 2.3$\pm$0.6 & 0.14$\pm$0.04 \\
Q2206$-$199\phantom{0a} & $<$43 & 7.0$\pm$1.3 & 0.05$\pm$0.04 \\
Q2343$+$125 & 38.8$\pm$9.0 & 6.2$\pm$1.1 & 0.05$\pm$0.04 \\
\hline
\end{tabular}
\begin{tablenotes}
\item $^{\rm a}$ Q0142$-$100 is gravitationally lensed by a factor of 3.2. The continuum flux densities have been corrected for magnification effects by dividing by 3.2.
\end{tablenotes}
\end{threeparttable}
\end{table}

\subsection{Line characterization}\label{lines}

\begin{figure*}
\centering
\makebox[\textwidth][c]{
\includegraphics[width=1.1\textwidth]{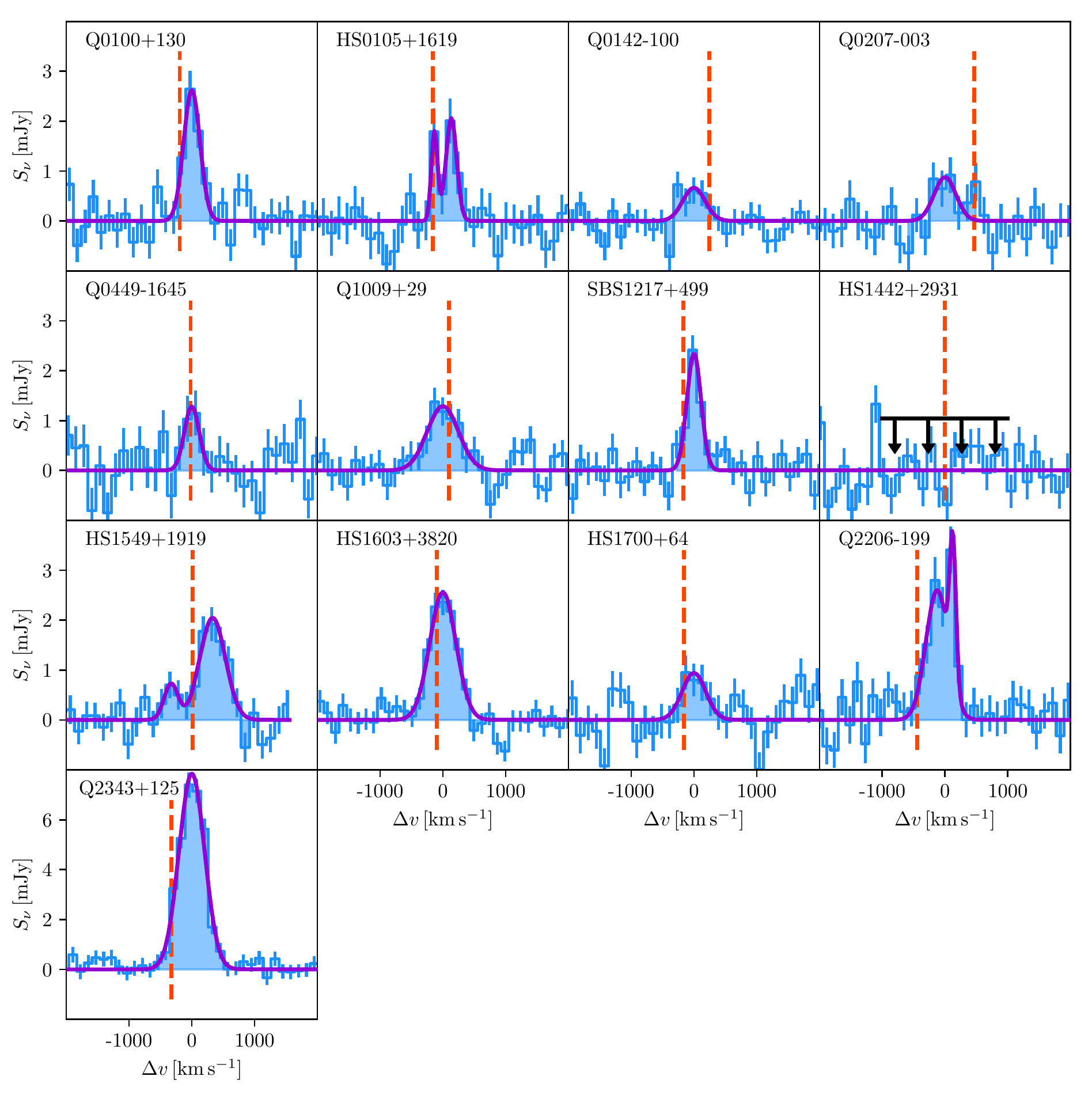}}
\caption{Continuum-subtracted CO(3--2) spectra extracted from the QSO positions. Single or double Gaussian models are shown as purple curves fit to the data, where the model with the highest likelihood (derived from Eq.~\ref{likelihood}) was selected. Each spectrum has been binned by a factor of 2. The shaded regions indicate the range over which we integrated to calculate the line luminosities. The dashed vertical red lines show the line centres based on the redshifts found by \citet{trainor2012} from near-IR lines. We did not detect the CO(3--2) line in HS1442$+$2931, so we show the 3$\sigma$ upper limit as a solid black line. Note the scale difference for the bright Q2343$+$125 CO source. Also note that Q0142$-$100 is gravitationally lensed by a factor of 3.2, and the spectrum shown has not been corrected for this.}
\label{fig:spectra_qso}
\end{figure*}

\begin{figure*}
\centering
\makebox[\textwidth][c]{
\includegraphics[width=1.1\textwidth]{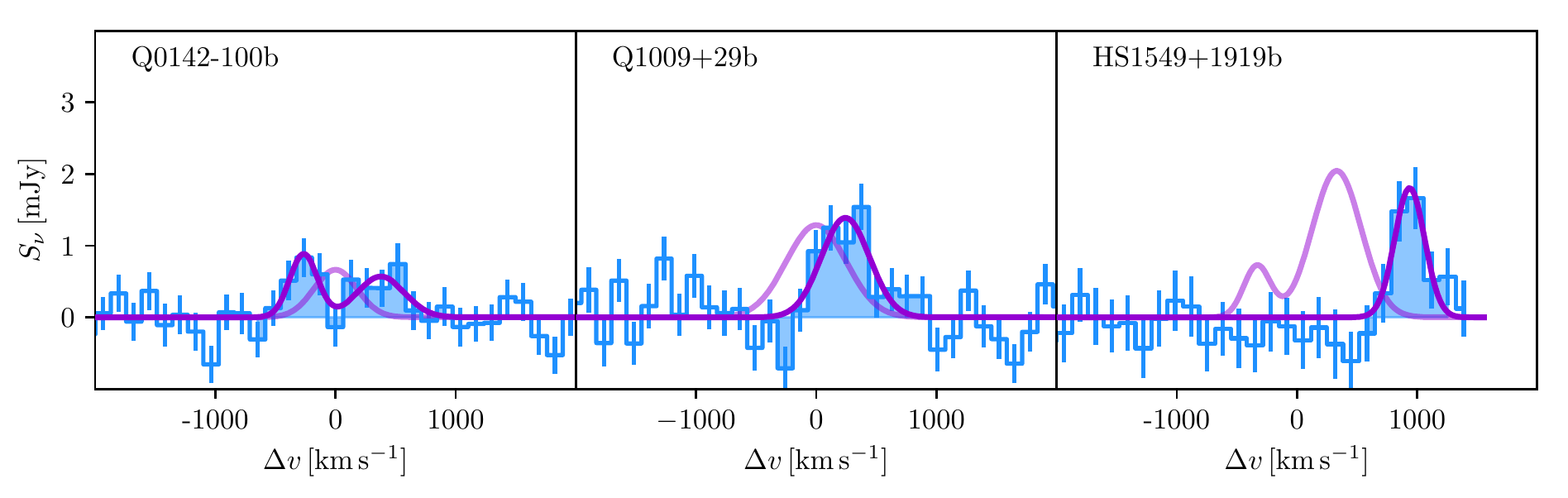}}
\caption{Extracted CO(3--2) spectra from the QSO companions, centred on the host QSO velocities. The purple curves show our best-fitting single and double Gaussian models, and the shaded regions are 3 times the best-fitting standard deviations, where we integrate to obtain the line luminosities. The faded purple curves are the best-fitting spectra of the host QSOs for reference.}
\label{fig:spectra_companion}
\end{figure*}

Looking at the spectra in Figs.~\ref{fig:spectra_qso} and \ref{fig:spectra_companion}, we can see that some sources show very clear double peaks that appear well described by a sum of two Gaussians (particularly HS0105$+$1619 and HS1549$+$1919). However, in some cases it is not so clear whether a single peak or a double peak best describes the data (e.g. Q2206$-$199), or even that the line feature is real. We therefore fit three models to our spectra: a constant; a single Gaussian; and a double Gaussian. We calculated and compared the resulting likelihoods of these models. The likelihoods, $\mathcal{L}$, were modelled by approximating the global minimum of $\chi^2$ as quadratic, which gives
\begin{equation}\label{likelihood}
\mathcal{L}=\left[\prod_{k=1}^M \Delta_k \right] \left[ \prod_{i=1}^N \left(2 \uppi \sigma_i^2\right)^{-1/2} \right] {\rm e}^{-\chi^2/2} |2 \uppi V|^{1/2}.
\end{equation}
\noindent
Here $\Delta_k$ is the prior for parameter $k$ (so if $x_k$ is the value of parameter $k$, and $x_{k, {\rm min}}$ and $x_{k, {\rm max}}$ are the minimum and maximum values of the priors, respectively, then $\Delta_k=[x_{k, {\rm max}} - x_{k, {\rm min}}]^{-1}$ for uniform priors and $\Delta_k=[x_k \ln (x_{k, {\rm max}}/x_{k, {\rm min}})]^{-1}$ for Jeffreys priors), $M$ is the number of fit parameters, $N$ is the number of data points, $\chi^2$ is the chi-squared of the best fit, and $V$ is the covariance matrix of the best fit. For the constant model, we used a uniform prior in the range $-$0.5 to 0.5\,mJy since the spectra have already been continuum-subtracted. In the single and double Gaussian models, for the amplitudes, we used a Jeffreys prior between 0.4 and 8\,mJy (0.4\,mJy being a typical value of three times the noise in the binned data and 8\,mJy being significantly higher than the largest peak observed); for the means, we used a uniform prior between $-$500 to 500\,km\,s$^{-1}$ relative to the previously-derived redshifts from \citealt{trainor2012} as this was around three times the uncertainties they reported; and for the FWHM, we used a Jeffreys prior between 50 to 800\,km\,s$^{-1}$ (the lower limit being about the width of one spectral bin and the upper limit being much larger than the widest line observed). For the fitting range, we used $\pm$2000\,km\,s$^{-1}$ relative to the peak of the observed line features.

The best fits are shown along with the raw spectra in Fig.~\ref{fig:spectra_qso} for the QSOs, and in Fig.~\ref{fig:spectra_companion} for the companions (note that the amplitude of the spectrum and corresponding fit shown for Q0142$-$100 has not been corrected for gravitational lensing). In Fig.~\ref{fig:spectra_qso} we also show for comparison the positions of the redshifts obtained by \citet{trainor2012} from broad near-IR hydrogen Balmer and [Mg{\sc II}] lines. We found that the likelihood function for three QSOs and one companion is maximized by a double Gaussian. 

To assess the significance of the detections we computed the logarithmic ratio of the likelihood function (Eq.~\ref{likelihood}) of the single or double Gaussian model (whichever provided a larger value, denoted as $\mathcal{L}_{\rm line}$) to that of the constant model ($\mathcal{L}_0$):
\begin{equation}\label{Q}
Q=\ln \left( \frac{\mathcal{L}_{\rm line}}{\mathcal{L}_0} \right).
\end{equation}
\noindent
The values of $Q$ are provided in Table \ref{table:qso}. As noted in Section \ref{reduction}, the significance of the weaker peaks were also assessed directly from the channel maps averaged over the width of the best-fitting lines, and we found them all to be $>3\sigma$ except for Q0142$-$100, which was found to be 2$\sigma$; we thus classify this source as a tentative detection.

From our best-fitting line profiles we extracted redshifts for each source, as well as as the amplitude and FWHM of the contributing lines. Line strengths were measured directly from the raw 1D spectra by integrating the flux densities in the range $[-3\sigma,3\sigma]$, where the $\sigma$ value was determined from the best fit, and centred on the best-fitting peak. In the case of a two-peaked fit, the integration bound was $[-3\sigma_{\rm L},3\sigma_{\rm R}]$, where $\sigma_{\rm L}$ and $\sigma_{\rm R}$ are from the left and right Gaussian fits, respectively. We note that deriving line strengths directly from the fits gave answers that differed by on average 9\,per cent. Luminosities were then derived using the luminosity distances.

Table \ref{table:qso} summarizes the CO(3--2) line properties of our sample. We report the positions (derived from the peaks of the CO(3--2) emission centroids, or for sources lacking a line, from the peaks of the continuum emission centroids), amplitudes, FWHMs, redshifts, line strengths, and luminosities from these measurements, for both the QSOs and the companions. For Q0142$-$100, the gravitationally lensed sourse, we provide the un-corrected best-fitting amplitude, but the line strength and luminosity have been corrected by the lensing factor of 3.2.

\setlength\tabcolsep{4pt}
\begin{table*}
\centering
\caption{Table of QSO CO(3--2) line properties, with companions in italics and given a suffix `b'. For sources whose spectra are best-fit by a sum of two Gaussians, we provide the amplitude, FWHM, and redshift of both lines. For sources where we do not detect a CO(3--2) line we report 3$\sigma$ upper limits on the line strenghts. The significance of the CO(3--2) line associated with Q0142$-$100 is 2$\sigma$, and we consider this to be a tentative detection.}
\label{table:qso}
\makebox[\textwidth][c]{
\begin{threeparttable}
\begin{tabular}{lcccccccc}
\hline
Name & RA/Dec$^{\rm a}$ & $Q^{\rm b}$ & CO(3--2) amplitude$^{\rm c}$ & CO(3--2) FWHM$^{\rm c}$ & $z_{\rm CO(3-2)}^{\rm d}$ & $F_{\rm CO(3-2)}^{\rm e}$ & ${\rm L}_{\rm CO(3-2)}$ & ${\rm L}^{\prime \ {\rm f}}_{\rm CO(1-0)}$ \\
 & [J2000] & & [mJy] & [km\,s$^{-1}$] & & [Jy\,km\,s$^{-1}$]  & [L$_{\odot} \times 10^{7}$] & [K\,km\,s$^{-1}$\,pc$^2 \times 10^{10}$] \\
 \hline
Q0100$+$130 & 01:03:11.27	$+$13:16:17.5 & 37.2 & 2.6$\pm$0.3 & 310$\pm$40 & 2.723 & 1.1$\pm$0.1 & 5.5$\pm$0.6 & \phantom{0}4.3$\pm$1.0 \\
HS0105$+$1619 & 01:08:06.40 $+$16:35:50.0 & 14.6 & 1.8$\pm$2.3/2.1$\pm$1.3 & 120$\pm$330/200$\pm$370 & 2.652/2.656 & 0.8$\pm$0.1 & 4.0$\pm$0.7 & \phantom{0}3.2$\pm$0.8 \\
\textit{HS0105$+$1619b} & 01:08:04.65 $+$16:35:45.1 & \dots & \dots & \dots & \dots & $<$1.4$^{\rm i}$ & \dots & \dots \\
Q0142$-$100 & 01:45:16.60 $-$09:45:17.0 & 5.2 & 0.7$\pm$0.2$^{\rm g}$ & 420$\pm$150$^{\rm g}$  & 2.740 & 0.07$\pm$0.03$^{\rm h}$ & 0.3$\pm$0.2$^{\rm h}$ & \phantom{0}0.3$\pm$0.1$^{\rm h}$ \\ 
\textit{Q0142$-$100b} & 01:45:16.74 $-$09:45:23.3 & 4.8 & 0.9$\pm$0.3/0.6$\pm$0.2 & 270$\pm$100/460$\pm$200 & 2.737/2.745 & 0.6$\pm$0.1 & 3.2$\pm$0.6 & \phantom{0}2.5$\pm$0.7 \\ 
Q0207$-$003 & 02:09:50.66 $-$00:05:05.8 & 4.5 & 0.9$\pm$0.3 & 410$\pm$160 & 2.866 & 0.6$\pm$0.1 & 3.2$\pm$0.8 & \phantom{0}2.5$\pm$0.8 \\
Q0449$-$1645 & 04:52:14.30 $-$16:40:17.6 & 1.2 & 1.3$\pm$0.4 & 280$\pm$110 & 2.684 & 0.3$\pm$0.1 & 1.7$\pm$0.7 & \phantom{0}1.3$\pm$0.6 \\
Q1009$+$29 & 10:11:55.60 $+$29:41:41.7 & 22.8 & 1.3$\pm$0.2 & 590$\pm$100 & 2.651 & 0.8$\pm$0.2 & 4.9$\pm$0.6 & \phantom{0}3.8$\pm$0.9 \\
\textit{Q1009$+$29b} & 10:11:55.06 $+$29:41:41.0 & 19.6 & 1.4$\pm$0.2 & 470$\pm$90 & 2.654 & 0.8$\pm$0.1 & 3.8$\pm$0.6 & \phantom{0}2.9$\pm$0.7 \\
SBS1217$+$499\phantom{0a} & 12:19:30.78 $+$49:40:52.6 & 38.0 & 2.3$\pm$0.3 & 270$\pm$40 & 2.706 & 0.9$\pm$0.1 & 4.1$\pm$0.5 & \phantom{0}3.4$\pm$0.8 \\
HS1442$+$2931 & 14:44:53.56 $+$29:19:05.6 & $-$1.3 & \dots & \dots & \dots & $<$0.5$^{\rm i}$ & $<$2.7$^{\rm j}$ & $<$2.1$^{\rm j}$ \\ 
HS1549$+$1919 & 15:51:52.45 $+$19:11:03.6 & 40.4 & 0.7$\pm$0.3/2.0$\pm$0.2 & 260$\pm$120/480$\pm$60 & 2.839/2.847 & 1.3$\pm$0.1 & 7.5$\pm$0.8 & \phantom{0}5.9$\pm$1.3 \\
\textit{Q1549$+$1919b} & 15:51:53.74 $+$19:11:08.5 & 10.0 & 1.8$\pm$0.4 & 290$\pm$70 & 2.855 & 0.7$\pm$0.1 & 4.1$\pm$0.5 & \phantom{0}3.2$\pm$0.7 \\
HS1603$+$3820\phantom{a} & 16:04:55.38 $+$38:12:01.8 & 127.3 & 2.6$\pm$0.2 & 490$\pm$40 & 2.552 & 1.5$\pm$0.1 & 7.1$\pm$0.4 & \phantom{0}5.5$\pm$1.1 \\
HS1700$+$64\phantom{00a} & 17:01:00.49 $+$64:12:09.4 & 3.9 & 0.9$\pm$0.3 & 450$\pm$140 & 2.753 & 0.6$\pm$0.1 & 3.1$\pm$0.7 & \phantom{0}2.4$\pm$0.7 \\
Q2206$-$199\phantom{0a} & 22:08:52.10 $-$19:43:59.0 & 49.6 & 2.6$\pm$0.6/2.8$\pm$1.4 & 420$\pm$120/120$\pm$470 & 2.577/2.580 & 1.6$\pm$0.1 & 7.4$\pm$0.7 & \phantom{0}5.7$\pm$1.2 \\
Q2343$+$125 & 23:46:28.25 $+$12:48:57.8 & 364.4 & 7.8$\pm$0.3 & 480$\pm$20 & 2.577 & 4.1$\pm$0.2 & 19.4$\pm$0.7 & 15.1$\pm$3.0 \\
\hline
\end{tabular}
\begin{tablenotes}
\item $^{\rm a}$ Coordinates of the centroid peak in the averaged CO(3--2) maps, except for HS0105$+$1619a, HS1442$+$2931, and Q2343$+$125a, where the coordinates are of the centroid peak in the averaged continuum maps.
\item $^{\rm b}$ Logarithmic ratio of the likelihood function of the single or double Gaussian model (whichever provided the larger value) to that of the constant model; see Eq.~\ref{Q}.
\item $^{\rm c}$ Parameters for the best-fitting single or double Gaussian models for the CO(3--2) emission lines.
\item $^{\rm d}$ Uncertainty in our CO redshift estimates are about 1\,per cent, which is the uncertainty obtained from the covariance matrices of the fit parameters.
\item $^{\rm e}$ Line strength, obtained by integrating the spectra from $[-3\sigma,3\sigma]$, where the $\sigma$ was determined from the best-fit, and centred on the best-fitting peak. In the case of a two-peaked fit, the integration bound was $[-3\sigma_{\rm L},3\sigma_{\rm R}]$, where $\sigma_{\rm L}$ and $\sigma_{\rm R}$ are from the left and right Gaussian fits, respectively.
\item $^{\rm f}$ $L^{\prime}_{\rm CO(3-2)}=\frac{c^2}{2 k_{\rm B}} \nu_{\rm obs}^{-2} D_{\rm L}^2 F_{\rm CO(3-2)} (1+z)^{-3}$, where $c$ is the speed of light, $k_{\rm B}$ is the Boltzmann constant, $\nu_{\rm obs}$ is the observed peak frequency, $D_{\rm L}$ is the luminosity distance in units of parsecs, $F_{\rm CO(3-2)}$ is the velocity-integrated line strength in units of W\,m$^{-2}$\,Hz$^{-1}$\,km\,s$^{-1}$, and we adopt the conversion factor $L^{\prime}_{\rm CO(3-2)}/L^{\prime}_{\rm CO(1-0)}=0.97\pm0.19$ \citep{carilli2013}.
\item $^{\rm g}$ Q0142$-$100 is gravitationally lensed by a factor of 3.2; we provide the best-fitting parameters of the un-corrected CO(3--2) line profile.
\item $^{\rm h}$ Q0142$-$100 is gravitationally lensed by a factor of 3.2; the line flux and line luminosity have been corrected for magnification effects by dividing by 3.2.
\item $^{\rm i}$ Assuming a characteristic FWHM of 350\,km\,s$^{-1}$ and taking the amplitude to be the 3$\sigma$ upper limit.
\item $^{\rm j}$ Assuming the redshift from \citet{trainor2012}.
\end{tablenotes}
\end{threeparttable}
}
\end{table*}

\section{Results}\label{results}

\subsection{Molecular gas redshifts}

Here we compare the redshifts of the submm emission lines to those from optical/UV observations. The QSO redshifts were measured from rest-frame far-UV lines \citep[such as N{\sc V}, C{\sc IV}, Si{\sc IV}, and C{\sc III}, see][]{rudie2012}, and more recently from near-IR Balmer lines \citep{trainor2012}. The far-UV-derived redshifts were found to be significantly blueshifted compared to those from the near-IR. The issue here is that the gas emitting the far-UV radiation is subject to different physical conditions compared to the gas emitting in the near-IR \citep[e.g.][]{richards2002}; lines in the near-IR are expected to do a better job at determining the systemic redshift. Our (rest frame) far-IR observations directly trace star-forming regions, and so it is useful to check if any differences show up statistically between the redshifts. 

We therefore compared our redshift measurements for the 12 cases where we have unambiguously detected the CO(3--2) line from the QSO to those found in the near-IR by \citet{trainor2012}. In the instances where a QSO was best-fit by two Gaussians, we used the flux-weighted mean of the two redshifts. We found the weighted mean and standard deviation of the differences to be $-$31$\,{\pm}\,$250\,km\,s$^{-1}$, which is consistent with no shift. Thus there appear to be no systematic velocity offsets between our measurements and those previously obtained from other lines. As a further test, we also looked for trends in the redshift differences as functions of CO(3--2) line luminosity, FIR luminosity, and AGN luminosity (see Sections \ref{FIRluminosity} and \ref{AGNluminosity}, respectively, for how the AGN luminosities and FIR luminosities were calculated), but these plots were consistent with random scatter given the large uncertainties.


\subsection{Resolving CO(3--2) emission regions}

\begin{figure}
\centering
\includegraphics[width=\columnwidth]{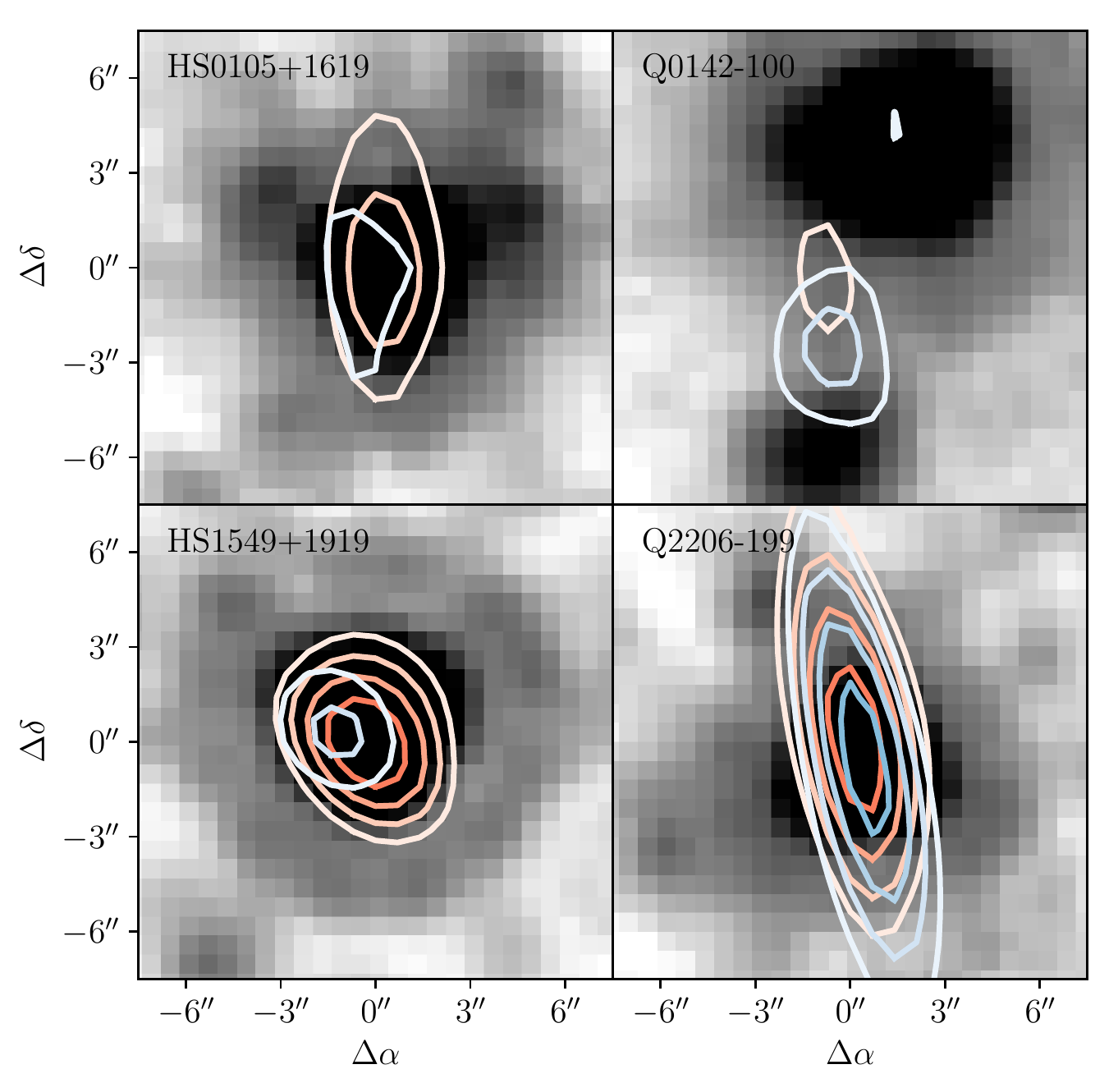}
\caption{15\,arcsec dirty-beam maps of the QSO regions where two lines were detected, plotted over IRAC 3.6-$\mu$m imaging. The red contours show the higher redshift peak, and the blue contours show the lower redshift peak. Contours here start at 3$\sigma_{\rm peak}$ and increase in steps of 1$\sigma_{\rm peak}$, where $\sigma_{\rm peak}$ is the standard deviation of the frequency bin at the location of the CO(3--2) peak being shown.}
\label{fig:maps_zoom}
\end{figure}

For cases where two peaks are detected spectrally, there could be additional information by looking at the two peaks spatially if the noise is small enough. To investigate this further, we produced enlarged 15\,arcsec by 15\,arcsec maps of the central QSO emission regions with contours centred on the frequency bins of the peaks (Fig.~\ref{fig:maps_zoom}). We also include Q0142$-$1000 here, since the companion is quite close to the host QSO and shows two CO(3--2) peaks (an indication that there could be some interesting interaction activity). In this figure, the contours start at 3$\sigma_{\rm peak}$ and increase in steps of 1$\sigma_{\rm peak}$, where $\sigma_{\rm peak}$ is the standard deviation of the stacked images around the frequency of the CO(3--2) peak in question.

We can see that in some sources the two emission regions are separated by 1--3\,arcsec (corresponding to 8--25\,kpc projected on the sky at $z=2.7$, e.g. HS1549$+$1919), and in some sources the two emission regions overlap (e.g., HS0105$+$1619 and Q2206$-$199). However, it is important to keep in mind that our observations have a synthesized resolution of around 3.7\,arcsec and that most of the fainter peaks are only detected at an S/N of 3, so some of the offsets might not be statistically significant.

For the simple case of perfectly circular and Gaussian beams (which is certainly not the case here), the accuracy with which a given source can be localized is related to the S/N value of the detection and the beamsize \citep[see e.g.][]{ivison2007}. Since the probability density of being separated by a distance $r$ in two dimensions is proportional to $r e^{-r^2/2 \sigma^2}$, it is then straightforward to estimate the probability of finding the given source separated by a distance greater than $r$ and hence assess whether or not it is reasonable that two nearby detections are the same source. In our case this rough calculation shows that the only significant (here meaning a probability less than 0.01) separation is that of Q0142$-$100a, where the two emission components are separated by 2.8\,arcsec and the positional uncertainty is 0.8\,arcsec. We also find that HS1549$+$1919 is potentially interesting, since the angular separation is 1.4\,arcsec, while the positional uncertainty is 0.6\,arcsec, but the probability of observing a separation larger than this is only at the 0.1 level. For the remaining sources with two emission peaks our data cannot rule out the possibility that we are seeing a single emission region.

\subsection{Submm-derived host galaxy properties}

\subsubsection{CO(1--0) luminosity and gas mass}

A quantity of interest to calculate is the CO(1--0) line luminosity, as this is expected to be closely related to the total gas mass of the host galaxy. Indeed, in the simplest case, the more CO gas a galaxy contains the brighter its CO lines will be, and under the assumption that the CO fraction of a galaxy's mass is relatively constant throughout a population, it can be easily seen how these two quantities can be correlated. The constant of proportionality between CO(1--0) line luminosity (in units of K\,km\,s$^{-1}$\,pc$^2$) and gas mass (in units of solar masses) is a quantity of great interest, and there have been numerous methods developed to determine its value for both local and diatant galaxies. For example, in the Milky Way the constant of proportionality is between 4 and 5\,M$_{\odot}$/(K\,km\,s$^{-1}$\,pc$^2$) \citep[e.g.,][]{solomon1991,liszt2010}, while for local IR galaxies it is probably much lower at around 0.5 to 1\,M$_{\odot}$/(K\,km\,s$^{-1}$\,pc$^2$) \citep[e.g.,][]{solomon2005,papadopoulos2012}, although some models predict values up to 5\,M$_{\odot}$/(K\,km\,s$^{-1}$\,pc$^2$); for more details see the review papers on molecular gas in galaxies by \citet{carilli2013} and \citet{bolatto2015}.

We calculated the CO(1--0) luminosities of our QSOs in two steps. First, we converted the line strength into units of $L^{\prime}_{\rm CO(3-2)}$, where
\begin{equation}
L^{\prime}_{\rm CO(3-2)}=\frac{c^2}{2 k_{\rm B}} \nu_{\rm obs}^{-2} D_{\rm L}^2 F_{\rm CO(3-2)} (1+z)^{-3}.
\end{equation}
\noindent
In this equation, $c$ is the speed of light, $k_{\rm B}$ is the Boltzmann constant, $\nu_{\rm obs}$ is the observed peak frequency, $D_{\rm L}$ is the luminosity distance in units of parsecs, and $F_{\rm CO(3-2)}$ is the velocity-integrated line strength in units of W\,m$^{-2}$\,Hz$^{-1}$\,km\,s$^{-1}$ \citep[see e.g.][]{solomon1997}. Second, we estimated the CO(1--0) luminosity using an estimate of the ratio $L^{\prime}_{\rm CO(3-2)}/L^{\prime}_{\rm CO(1-0)}=0.97\pm0.19$, appropriate for QSOs, taken from \citet{carilli2013}. For this conversion factor we adopted an uncertainty of 20\,per cent, the typical spread from their sample. For completeness, we also calculated our line luminosities in solar units using
\begin{equation}
L_{\rm CO(3-2)} = 4 \uppi D_{\rm L}^2 F_{\rm CO(3-2)}/{\rm L}_{\odot},
\end{equation}
where $D_{\rm L}$ is the luminosity distance, now in units of m, and $F_{\rm CO(3-2)}$ is this time the frequency-integrated line strength in units of W\,m$^{-2}$ \citep[see e.g.][]{marsden2005}.

Measurements for each source are given in Table \ref{table:qso}. We also give gas mass measurements in Table \ref{table:properties} by adopting a conversion factor of 1\,M$_{\odot}$/(K\,km\,s$^{-1}$\,pc$^2$); there is clearly lots of uncertainty as to what this conversion factor should really be, and we have simply chosen unity for simplicity. We therefore caution the reader that the gas mass measurements we provide here should be interpreted somewhat loosely. For Q0142$-$100, we have corrected the luminosity for gravitational lensing by dividing by a factor of 3.2.

\subsubsection{Far-IR luminosity and dust mass}\label{FIRluminosity}

A useful quantity to compare to our measurements of $L^{\prime}_{\rm CO(1-0)}$ is the far-IR luminosity, $L_{\rm FIR}$, defined as the specific luminosity integrated from 42.5 to 122.5\,$\mu$m in the rest-frame. This quantity is effectively the total energy output of warm dust, which has been heated up by stars, making it a useful proxy for the star-formation rate of a galaxy \citep[e.g.][]{kennicutt98}. The far-IR luminosity of a galaxy is also is important for understanding the total amount of dust present in the galaxy, for the more dust there is the brighter the galaxy will be in the far-IR. There is a well-studied correlation between $L^{\prime}_{\rm CO(1-0)}$ and $L_{\rm FIR}$ \citep[see e.g.][]{carilli2013}, and the ratio $L^{\prime}_{\rm CO(1-0)}$/$L_{\rm FIR}$ is expected to be a good tracer of the rate at which gas is turned into stars (i.e.~the gas consumption time-scale, see \citealt{bolatto2015}); we would thus like to know where the hyper-luminous QSOs in our sample lie compared to other populations from the literature. We will compare with representative samples of low and high redshift Type I QSOs (i.e.~QSOs that are not dust-obscured and have bright rest-frame UV continua), low-$z$ Type II QSOs (i.e.~dust-obscured QSOs containing very faint UV continua), and SMGs; however, before we proceed, we need to understand how $L_{\rm FIR}$ has been estimated by the other authors to ensure that we are comparing similar quantities.

We will use low redshift ($z<0.4$) QSOs with CO observations from the Palomar-Green (PG) survey \citep{scoville2003,evans2006}, the Hamburg-ESO (HE) survey \citep{bertram2007}, and a sample that was selected from several IR surveys \citep{xia2012}. For these QSOs, \citet{xia2012} provide $L_{\rm FIR}$ estimates based on 60 and 100\,$\mu$m IRAS photometry using the relation $F_{\rm FIR}=1.26 \times 10^{-14} \left( 2.58 S_{60}+1.00 S_{100} \right)$, where $F_{\rm FIR}$ is the flux in units of W\,m$^{-2}$ (so knowledge of the redshfit/distance is used to obtain a luminosity) and $S_{60}$ and $S_{100}$ are the flux densities at 60 and 100\,$\mu$m, respectively, in units of Jy \citep[see][]{helou1988}. 

The high redshift ($z>1.4$ and up to around 6) QSOs with CO observations we use are from \citet{walter2003}, \citet{solomon2005}, \citet{carilli2007}, \citet{maiolino2007}, \citet{coppin2008}, \citet{wang2010}, \citet{simpson2012}, and \citet{fan2018}. For these sources, $L_{\rm FIR}$ values come from \citet{wang2010}, \citet{xia2012}, and \citet{simpson2012} using mm and submm flux density measurements to normalize a modified blackbody function with a dust temperature of 47\,K (or 40\,K in the case of \citealt{simpson2012}) and an emissivity index of 1.6; this choice of dust temperature and emissivity index are consistent with previous modified blackbody fits of high redshift QSOs from e.g., \citet{beelen2006,wang2008}.

The low redshift Type II QSOs come from various {\it Spitzer\/} far-IR surveys with follow-up submm/mm observations. We use five CO detections of Type II QSOs from \citet{krips2012}, five from \citet{villar-martin2013}, and one from \citet{rodriguez2014}. In \citet{krips2012}, $L_{\rm FIR}$ was estimated using {\it Spitzer\/} far-IR photometry using the equation $L_{\rm FIR}=0.82 L_{70}+1.35 L_{160}$, where $L_{70}$ and $L_{160}$ are the monochromatic luminosities at 70 and 160\,$\mu$m, respectively \citep[see][]{dale2002}, while in \citet{villar-martin2013} the authors used {\it WISE} and {\it Spitzer\/} photometry to fit spectral energy distributions (SEDs) directly. Lastly, \citet{rodriguez2014} used the same far-IR-based equation that was applied to the low-redshift sample studied by \citet{xia2012}.

The SMGs we use are from \citet{ivison2011}, \citet{bothwell2013}, and \citet{aravena2016}. Specifically, \citet{bothwell2013} calculates far-IR luminosities using the far-IR-radio correlation via $L_{\rm FIR}=4\uppi D_{\rm L}^2 (8.4 \times 10^{14}) S_{1.4\,\mathrm{GHz}}(1+z)^{\alpha-1} {\rm L}_{\odot}$, where $S_{1.4\,\mathrm{GHz}}$ is the 1.4-GHz flux density and $\alpha=0.8$ \citep{yun2001}. The luminosities in \citet{ivison2011} and \citet{aravena2016} are based on direct far-IR photometry and fitting SEDs. The SMGs from \citet{aravena2016} are gravitationally lensed, and we have corrected their line and far-IR luminosities using the magnification strengths provided in the paper.

From Ross et al.~(in preparation), we have at our disposal 850-$\mu$m flux densities for every source in our sample and 450-$\mu$m flux densities for some, and we have also measured 3-mm continuum flux densities from NOEMA (see Table \ref{table:continuum}). These data match most closely to the high redshift QSOs, so in order to compare our sample to those from the literature as closely as possible, we adopt the same technique of fitting a modified blackbody with a dust temperature of 47\,K and an emissivity index of 1.6, which again are consistent with modified blackbody fits to other high redshift QSOs \citep[e.g.,][]{beelen2006,wang2008}. The uncertainty was estimated by combining in quadrature the uncertainty from the fit and an additional 50\,per cent, which we found to be the resulting spread from varying the dust temperature by $\pm10$\,K. For Q0142$-$100, we use the de-magnified flux densities given in Table \ref{table:qso}. The results are provided in in Table \ref{table:properties}. We find that this model provides a good fit to all of our sources, and it allows us to compare our sample more easily to the other sources from the literature (Ross et al.~in preparation fit the dust temperatures directly and find values of around 50\,K). The only exception is HS1442$+$2931, where the 3\,mm flux density is much higher than predicted by a modified blackbody function, and is likely not thermal, so we exclude this measurement from the fit.

\begin{figure*}
\centering
\includegraphics[width=\columnwidth]{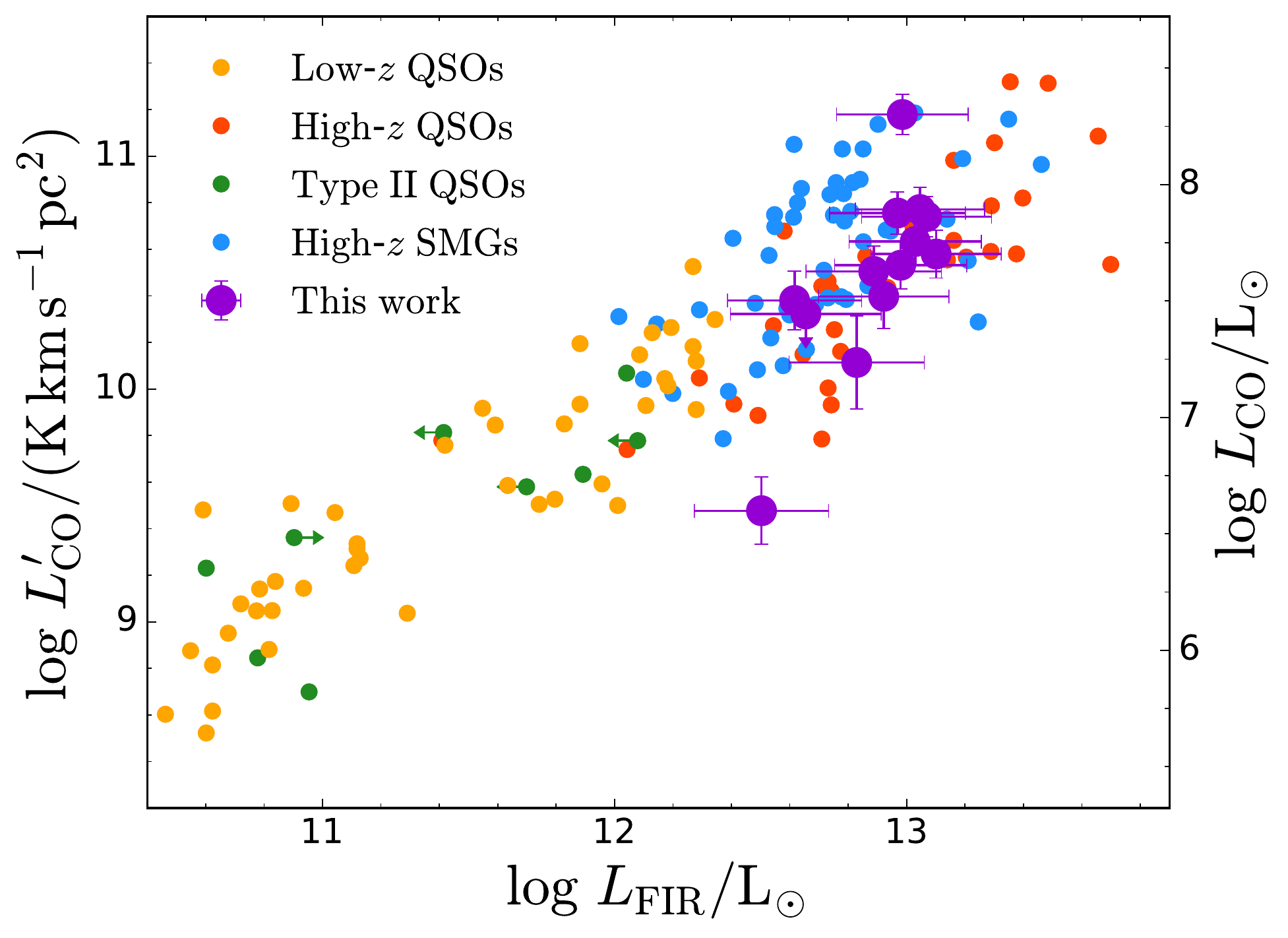}
\includegraphics[width=\columnwidth]{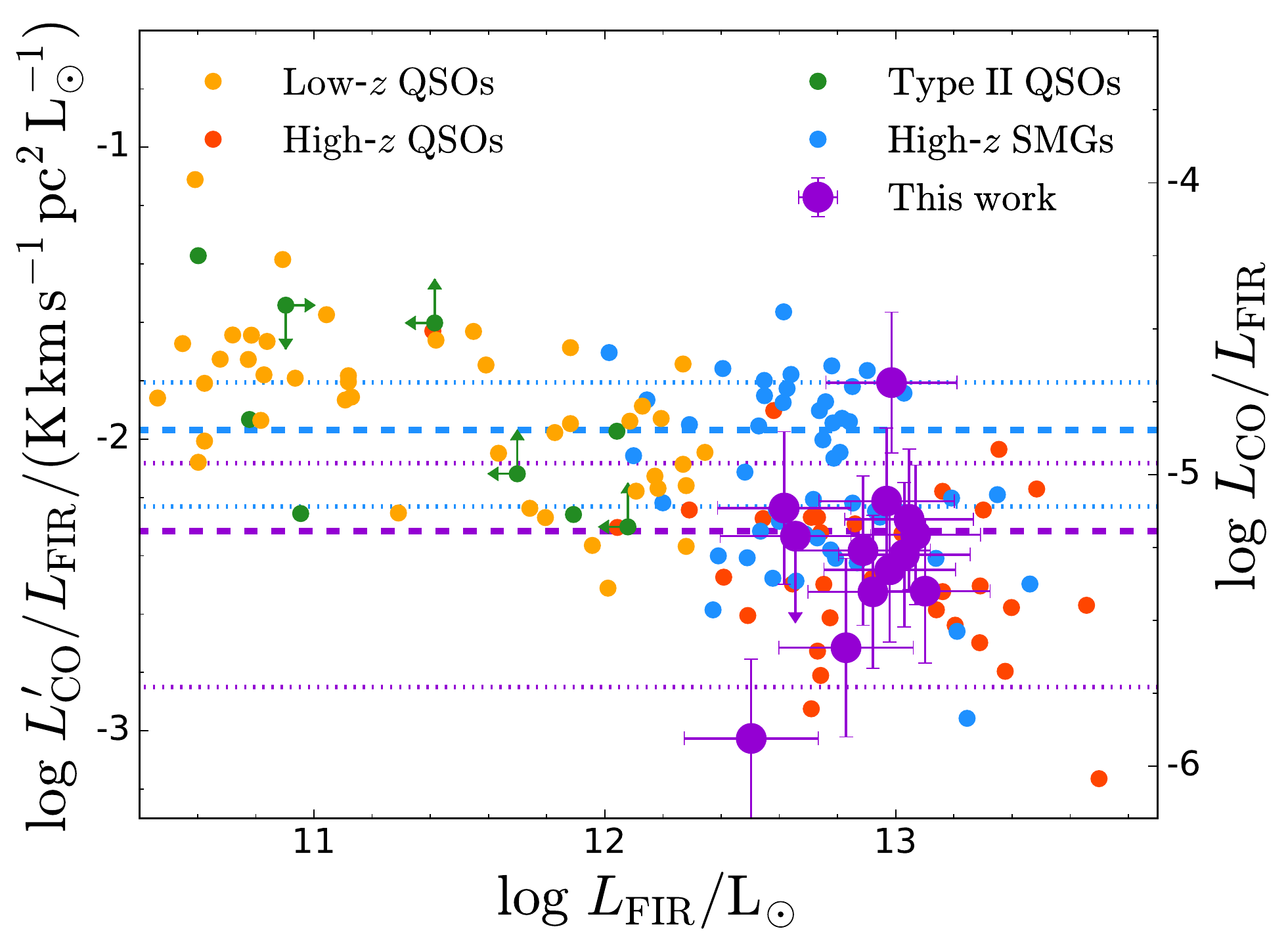}
\caption{{\it Left:} Relation between $L^{\prime}_{\rm CO(1-0)}$ and ${\rm L}_{\rm IR}$. We compare our sample to low-$z$ QSOs, high-$z$ QSOs, low-$z$ Type II QSOs, and SMGs (references can be found in the text). Our sample lies at the high end of the $L^{\prime}_{\rm CO(1-0)}$--${\rm L}_{\rm IR}$ relation, and seems to fit the trend of high-$z$ QSOs within the scatter. For comparison, we also show the $y$-axis scaled to units of $L_{\rm CO(1-0)}$. {\it Right:} The ratio $L^{\prime}_{\rm CO(1-0)}/L_{\rm FIR}$, which is a proxy for the gas consumption time-scale since $L^{\prime}_{\rm CO(1-0)}$ is proportional to the gas mass and $L_{\rm FIR}$ is proportional to the star-formation rate. The mean gas consumption time-scales for our hyper-luminous QSOs and SMGs are shown as horizontal dashed curves, and standard deviations are shown as horiontal dotted curves. The ratio of the hyper-luminous QSO to SMG mean gas consumption time-scale is 0.45$\pm$0.38. For comparison, we also show the $y$-axis scaled to units of $L_{\rm CO(1-0)}/L_{\rm FIR}$.}
\label{luminosity}
\end{figure*}

Figure \ref{luminosity} shows the resulting $L_{\rm FIR}$ values plotted against $L^{\prime}_{\rm CO(1-0)}$ and the ratio $L^{\prime}_{\rm CO(1-0)}/L_{\rm FIR}$ (for comparison, we also show the $y$-axis scaled to $L_{\rm CO(1-0)}$), along with low and high redshift QSOs and SMGs from the literature. We find that our sample is very similar to the high redshift QSOs, despite the fact that they have been selected as some of the brightest QSOs in the sky. Our sample also has comparable CO luminosities to SMGs but somewhat higher far-IR luminosities. This is most noticeable in the plot of $L^{\prime}_{\rm CO(1-0)}/L_{\rm FIR}$, in other words the gas consumption time-scale, where we see that the QSOs in our sample (and other high redshift QSOs from the literature) have on average smaller ratios (this observation is quantified and compared to similar results from the literature in Section \ref{discussion}). Q0142$-$100 is an outlier owing to the fact that it is not as intrinsically luminous a QSO.

The far-IR luminosities of the host galaxies are related to their dust masses through their dust temperatures, and have been estimated by Ross et al.~(in preparation) using measurements of the dust temperatures obtained from 450- and 850-$\mu$m SCUBA-2 observations (where only upper limits to the 450-$\mu$m flux density were available, a value of 40\,K was assumed). We provide these dust mass measurements in Table \ref{table:properties}, and compute the dust-to-gas ratio for each of our sources. We find that the median dust-to-gas ratio in our sample is 1.7$^{+0.8}_{-0.5}$\,per cent, where the uncertainty was estimated from bootstrap resampling. This ratio can be compared to what has been found in other high redshift QSOs and SMGs. For QSOs, \citet{schumacher2012} found a dust-to-gas ratio of 0.2--3\,percent in a $z=2.8$ dust-obscured QSO, \citet{wang2016} found this ratio to be 1.5\,percent in another hyper-luminous QSO at $z=6.3$, and \citet{michalowski2010} found ratios ranging from 1.5--4.3\,percent in a sample of nine QSOs ranging from $z=5$ to 6.5. For SMGs, a large {\it Herschel\/} survey found dust-to-gas ratios typically of order 1\,percent \citep{santini2010} with a spread of about 0.5\,dex, while \citet{swinbank2014} estimated a comparable value from a ground-based survey. Considering the uncertainties involved in computing such a quantity, and the spread within individual populations, we find no evidence that our sample of QSOs differs from typical QSOs and SMGs.

\subsubsection{FWHM and dynamical mass}

We would also like to compare our FWHM measurements to other QSOs and SMGs from the literature. The virial theorem tells us that the linewidths (squared) from our observations are proportional to the dynamical masses of the host galaxies and inversely proportional to their radii, which makes this an interesting quantity to explore in further detail. However, there are a couple of caveats: first, there is the unknown geometry of the system, which could be disk-like, spherical, or highly irregular due to for example a merger; and second, in the case of a disk, there is an unknown inclination angle so that what we actually measure is FWHM/$\sin i$. For the purposes of this analysis we will assume the galaxies hosting the observed QSOs to be disk-like, and incorporate the unknown inclination angle. Under this assumption, since $L^{\prime}_{\rm CO(1-0)}$ is proportional to the gas mass, the ratio $L^{\prime}_{\rm CO(1-0)} \sin^2 i /$FWHM$^2$ should also be proportional to the gas-mass fraction at a fixed host-galaxy radius.

\begin{figure}
\centering
\includegraphics[width=\columnwidth]{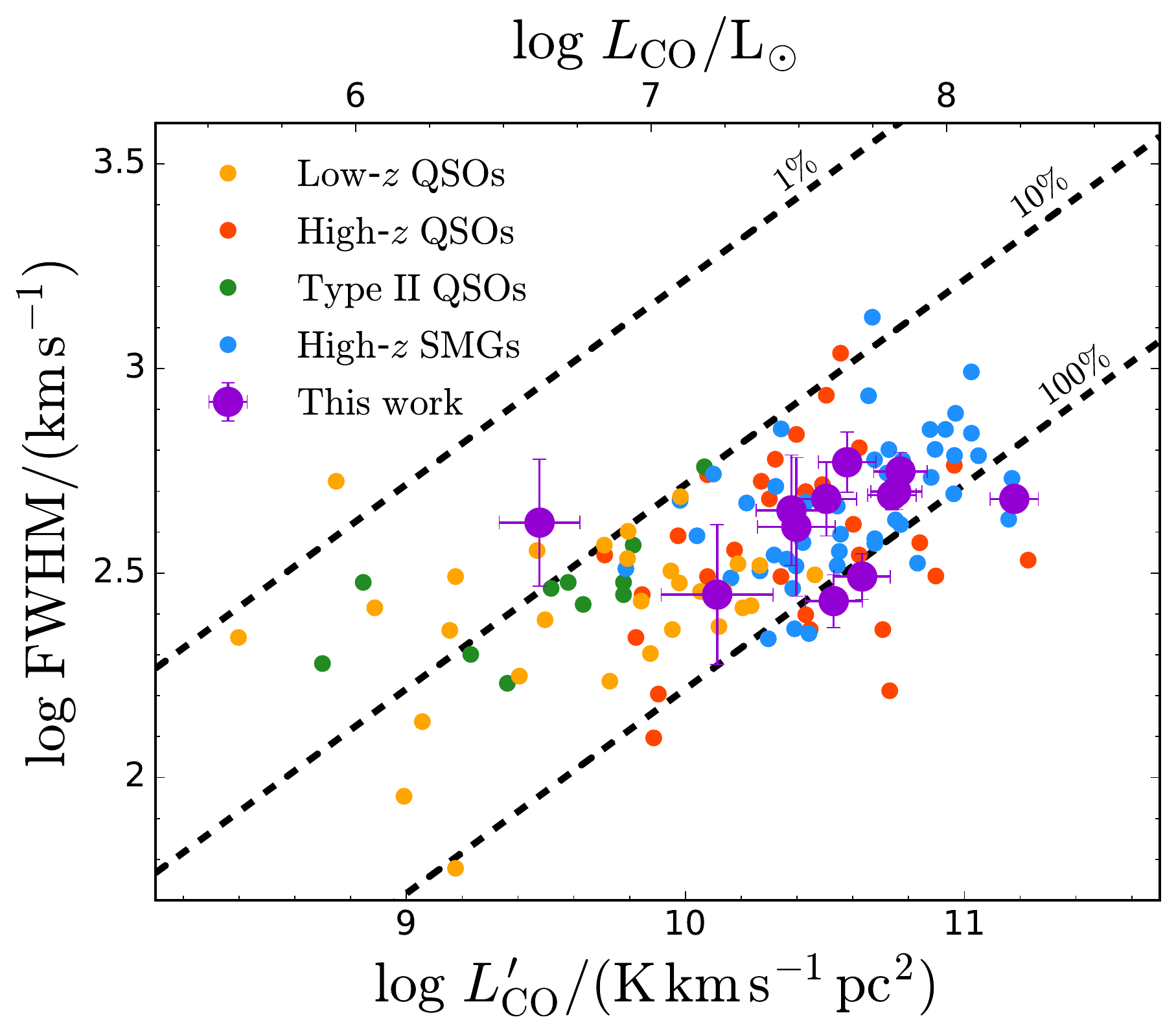}
\caption{Line FWHM versus $L^{\prime}_{\rm CO(1-0)}$ (and, for comparison, $L_{\rm CO(1-0)}$). Where our CO(3--2) profile is best-fit by two Gaussians we use a FWHM from fitting the entire profile by a single Gaussian model. We compare our measurements to low-$z$ QSOs, high-$z$ QSOs, low-$z$ Type II QSOs, and SMGs. Our QSOs fit within the scatter of the other classes of objects. We also show lines of constant gas-mass fraction, assuming an inclination angle of 32.7\,deg (the average inclination angle expected from an isotropic distribution) and galaxy radii of 10\,kpc (which is proportional to the ratio of $L^{\prime}_{\rm CO(1-0)} \sin^2 i /$FWHM$^2$).}
\label{fwhm}
\end{figure}

For the purpose of comparing single-peaked profiles to double-peaked (or even triple-peaked) profiles, it is common to fit a single Gaussian to all spectra and plot the resulting FWHM values. This was done for the sources in our sample with double-peaks, and in Fig.~\ref{fwhm} we show our results as a function of $L^{\prime}_{\rm CO(1-0)}$ and $L_{\rm CO(1-0)}$. Also shown are QSOs from the literature for which the FWHM values from a single Gaussian fit are available: for low-redshift QSOs we show values from \citet{evans2006} and \citet{xia2012}; for high-redhsift QSOs we show values from \citet{walter2003}, \citet{solomon2005}, \citet{carilli2007}, \citet{maiolino2007}, \citet{coppin2008}, \citet{wang2010}, and \citet{simpson2012}; for low-redshift Type II QSOs we show values from \citet{krips2012}, \citet{villar-martin2013}, and \citet{rodriguez2014}; and for SMGs we show values from \citet{ivison2011}, \citet{bothwell2013}, and \citet{aravena2016}.

In this comparison we have used linewidths from CO(3--2) transitions when possible. We note that some observations were of CO(1--0), CO(2--1), CO(5--4), and CO(6--5); however, this should not dramatically alter the linewidths, assuming that the emission is coming from the same molecular gas regions. This is typically not true for the higher transition lines, which require more energetic astrophysical processes to become excited, but for the lower transition lines here this assumption is reasonable.

In Fig.~\ref{fwhm} we also show lines of constant gas-mass fraction, assuming a mean velocity reduction factor of $\langle\sin i\rangle =0.79$ \citep[see Appendix A of][]{law2009} and a fiducial radius of 1\,kpc, motivated by resolved observations of molecular gas in strongly lensed QSOs \citep[e.g.,][]{alloin1997,gallerani2012,anh2013,leung2017}. In this plot, the gas-mass fractions are increasing towards the bottom right. We also provide quantitative values of the dynamical masses and gas-mass fractions in Table \ref{table:properties}, assuming the same velocity reduction factor of 0.79 and fiducial values of 1\,kpc for the radii containing all of the gas in the galaxies; we note that these dynamical masses are comparable to those seen in typical star-forming galaxies \citep[e.g.,][]{erb2003,price2016}. For the QSOs in our sample, the gas mass fractions span about 40 to 60\,percent, but we note that two QSOs, Q0100$+$130 and SBS1217$+$499, have gas mass fractions over 100\,percent (which is clearly unphysical), but the uncertainties overlap with numbers less than 100. Q2343$+$125, on the other hand, also has an estimated gas mass fraction above 100\,percent, with no such overlapping errorbars. This is almost certainly due to the very loose assumptions used throughout the calculation (i.e.~the CO line luminosity-to-gas mass factor, the inclination angle, the radius of the CO-emitting region), one or more of which were poor in the case of this source. Q0142$-$100 has an anomalously low gas-mass fraction, but being the only source that is gravitationally lensed this may not be surprising as emission regions of different physical scales will be magnified by different strengths, and this has not been taken into account in our analysis. 

For comparison with other populations from the literature, several surveys of SMGs have found gas-mass fractions that range from 20--80\,percent \citep{tacconi2006,tacconi2013,wiklind2014}, and a similar (but potentially smaller) range of 15--70\,percent has been seen in high-redshift QSOs \citep{maiolino2007,coppin2008,leung2017}; the gas-mass fractions found in our survey seem to be consistent with these published results. We note that a similar comparison between gas-mass fractions of high redshift QSOs and SMGs by \citet{simpson2012} found no evidence for a difference between the two classes of sources, although their sample contained only two QSOs, so their result was highly dependent on the assumed inclination angles. Our sample, containing 12 sources, still depends on the individual inclination angles, but adopting an average angle should provide a better idea of any differences (or lack thereof) between QSOs and SMGs. We find that the QSOs in our study lie within the spread of the other classes of sources from the literature, including SMGs, suggesting (as also found by \citealt{simpson2012}) that there is no significant difference in gas-mass fraction between the two populations. 

\begin{table*}
\centering
\caption{Table of derived QSO properties.}
\label{table:properties}
\begin{threeparttable}
\begin{tabular}{|lcccccccc|}
\hline
Name & $M_{\rm gas}^{\rm a}$ & $L_{\rm FIR}^{\rm b}$ & $M_{\rm dust}^{\rm c}$ & $M_{\rm dyn}^{\rm d}$ & $L_{\rm AGN}^{\rm e}$ & $M_{\rm BH}^{\rm f}$ & Gas-mass & Dust-to-gas \\
 & [M$_{\odot}\times10^{10}$] & L$_{\odot}\times10^{13}$ & [M$_{\odot}\times10^8$] & [M$_{\odot}\times10^{10}$] & L$_{\odot}\times10^{14}$ & M$_{\odot}\times10^{9}$ & fraction$^{\rm g}$ [\%] & ratio$^{\rm h}$ [\%] \\
 \hline
Q0100$+$130 & 4.3$\pm$1.0 & 1.2$\pm$0.6 & 5.2$\pm$1.0 & \phantom{1}3.6$\pm$0.9 & 3.6$\pm$1.6 & 2.0 & 121$\pm$41 & 1.2$\pm$0.4 \\
HS0105$+$1619 & 3.2$\pm$0.8 & 0.8$\pm$0.4 & 3.3$\pm$0.5 & \phantom{1}8.5$\pm$3.6 & 2.5$\pm$1.1 & 1.4 & \phantom{1}37$\pm$18 & 1.0$\pm$0.3 \\
Q0142$-$100 & 0.3$\pm$0.1$^{\rm i}$ & 0.3$\pm$0.2$^{\rm i}$ & 1.9$\pm$0.3$^{\rm i}$ & \phantom{1}6.5$\pm$4.7 & $<$1.1$^{\rm i}$ & $<$0.7$^{\rm i}$ & \phantom{1}5$\pm$4 & 6.3$\pm$2.3 \\ 
Q0207$-$003 & 2.5$\pm$0.8 & 0.8$\pm$0.4 & 6.1$\pm$0.9 & \phantom{1}6.2$\pm$4.9 & 3.4$\pm$1.5 & 1.9 & \phantom{1}40$\pm$34 & 2.4$\pm$0.9 \\
Q0449$-$1645 & 1.3$\pm$0.6 & 0.7$\pm$0.4 & 4.1$\pm$0.8 & \phantom{1}2.9$\pm$2.3 & 2.2$\pm$1.0 & 1.3 & \phantom{1}45$\pm$41 & 3.2$\pm$1.6 \\
Q1009$+$29 & 3.8$\pm$0.9 & 1.3$\pm$0.7 & 7.7$\pm$1.0 & 12.9$\pm$4.4 & 6.1$\pm$2.7 & 3.4 & \phantom{1}29$\pm$12 & 2.0$\pm$0.5 \\
SBS1217$+$499\phantom{0a} & 3.4$\pm$0.8 & 1.0$\pm$0.5 & 9.8$\pm$2.2 & \phantom{1}2.7$\pm$0.8 & 2.9$\pm$1.3 & 1.6 & 126$\pm$48 & 2.9$\pm$0.9 \\
HS1442$+$2931 & $<$2.1 & 0.5$\pm$0.3 & 2.4$\pm$0.9 & \dots & 2.7$\pm$1.2 & 1.5 & \dots & $>$1.1 \\ 
HS1549$+$1919 & 5.9$\pm$1.3 & 1.1$\pm$0.6 & 7.6$\pm$1.1 & 11.6$\pm$2.5 & 8.3$\pm$3.7 & 4.6 & \phantom{1}51$\pm$16 & 1.3$\pm$0.3 \\
HS1603$+$3820\phantom{a} & 5.5$\pm$1.1 & 1.2$\pm$0.6 & 7.5$\pm$1.2 & \phantom{1}8.9$\pm$1.5 & 6.1$\pm$2.7 & 3.4 & \phantom{1}62$\pm$16 & 1.4$\pm$0.3 \\
HS1700$+$64\phantom{00a} & 2.4$\pm$0.7 & 0.4$\pm$0.2 & 6.2$\pm$1.6 & \phantom{1}7.5$\pm$4.7 & 7.6$\pm$3.4 & 4.3 & \phantom{1}32$\pm$22 & 2.6$\pm$1.0 \\
Q2206$-$199\phantom{0a} & 5.7$\pm$1.2 & 0.9$\pm$0.5 & 6.1$\pm$1.1 & \phantom{1}9.3$\pm$1.9 & 2.5$\pm$1.1 & 1.4 & \phantom{1}62$\pm$18 & 1.1$\pm$0.3 \\
Q2343$+$125 & 15.1$\pm$3.0 & 1.0$\pm$0.5 & 8.9$\pm$1.6 & \phantom{1}8.5$\pm$0.7 & 2.1$\pm$0.9 & 1.2 & 177$\pm$38 & 0.6$\pm$0.1 \\
\hline
\end{tabular}
\begin{tablenotes}
\item $^{\rm a}$ Using a conversion factor of $L^{\prime}_{\rm CO(1-0)}$ to $M_{\rm gas}$ of 1 in these units \citep[e.g.,][]{solomon2005,liszt2010,papadopoulos2012,carilli2013,bolatto2015}.
\item $^{\rm b}$ Specific luminosity integrated from 42.5 to 122.5\,$\mu$m. SEDs were modeled as modified blackbody functions with a dust emissivity index of 1.6 and a temperature of 47\,K.
\item $^{\rm c}$ From Ross et al.~(in preparation).
\item $^{\rm d}$ Assuming a mean velocity reduction factor of $\langle\sin i\rangle =0.79$, appropriate for an isotropic distribution of disks, and a fiducial radius of 1\,kpc.
\item $^{\rm e}$ $L_{\rm AGN}=9.74 \left( 1450/4400 \right)^{\alpha+1} {\rm L}_{1450}$, where ${\rm L}_{1450}$ is the monochromatic luminosity at 1450\,\AA\ and $\alpha=-0.5$.
\item $^{\rm f}$ From \citet{trainor2012}; note that uncertainties are not provided in this paper.
\item $^{\rm g}$ The gas-mass fraction is defined as $M_{\rm gas}$/$M_{\rm dyn}$.
\item $^{\rm h}$ The dust-to-gas ratio is defined as $M_{\rm dust}$/$M_{\rm gas}$.
\item $^{\rm i}$ Corrected for gravitational lensing by dividing by a factor of 3.2 as these quantities depend on flux density amplitudes.
\end{tablenotes}
\end{threeparttable}
\end{table*}

\subsection{AGN luminosity and black hole mass}\label{AGNluminosity}

Another interesting quantity to calculate is the AGN bolometric luminosity, $L_{\rm AGN}$. The reason why the AGN bolometric luminosity is interesting is that it is basically a measure of the total light output of the accretion disk surrounding the central black hole (BH), thus it can be expected that the larger this luminosity, the larger the BH.

For the low- and high-redshift QSOs shown in Fig.~\ref{luminosity}, most of the values were taken directly from \citet{xia2012}, who calculated the bolometric luminosities using rest-frame $B$ band (4400\,\AA) monochromatic luminosities (i.e.~taking $\nu L_{\nu}$) with a conversion factor of 9.7$\pm$4.3. This conversion factor was taken from \citet{vestergaard2004}, who estimated the ratio of averaged QSO SEDs integrated over all wavelengths to the average flux density at the rest-frame wavelength of 4400\,\AA. However, in a few cases data were only available at 1450\,\AA, so \citet{xia2012} converted the monochromatic luminosities to 4400\,\AA\ assuming a power-law flux density distribution, $S_{\nu} \propto \nu^{\alpha}$, with a spectral index of $-0.5$ \citep[see e.g.,][]{schmidt1995,fan2001}. The three high-redshift QSOs from \citealt{fan2018} are not included in this compilation, and we cannot calculate their AGN luminosities because only IR rest-frame photometry is available. For the sample from \citet{simpson2012}, 1350\,\AA\ luminosities are available, which we correct to 1450\,\AA\ using the same prodecure outlined above. The low-redshift Type II QSOs are, by definition, too faint in the rest-frame UV to have an estimate of $L_{\rm AGN}$ and so cannot be included in this comparison.

For our sample, \citet{trainor2012} provide monochromatic luminosities at a rest-frame wavelength of 1450\,\AA. This means that in order to compare to the other sources from the literature properly, we must also follow the two-step conversion procedure, going to 4400\,\AA\ and then converting to bolometric luminosities. Specifically, we use the eqution
\begin{equation}\label{eq:agn}
L_{\rm AGN}=9.7 \left( 1450/4400 \right)^{\alpha+1} L_{1450},
\end{equation}
\noindent
where $L_{1450}$ is the monochromatic luminosity at 1450\,\AA\ and $\alpha=-0.5$. This gives us a final conversion factor of 5.6 (with an uncertainty of 2.5 propagated through from the uncertainty in the conversion factor from monochromatic luminosity to bolometric luminosity); we note that \citet{vestergaard2004} also provides a conversion factor of 4.7 to get from rest-frame 1450\,\AA\ monochromatic luminosity to bolometric luminosity, but this was not used by \citet{xia2012}, and the conversion factors are quite close in any case. We have divided the derived AGN luminosity by 3.2 for Q0142$-$100 to take the gravitational lensing into account.

\begin{figure}
\centering
\includegraphics[width=\columnwidth]{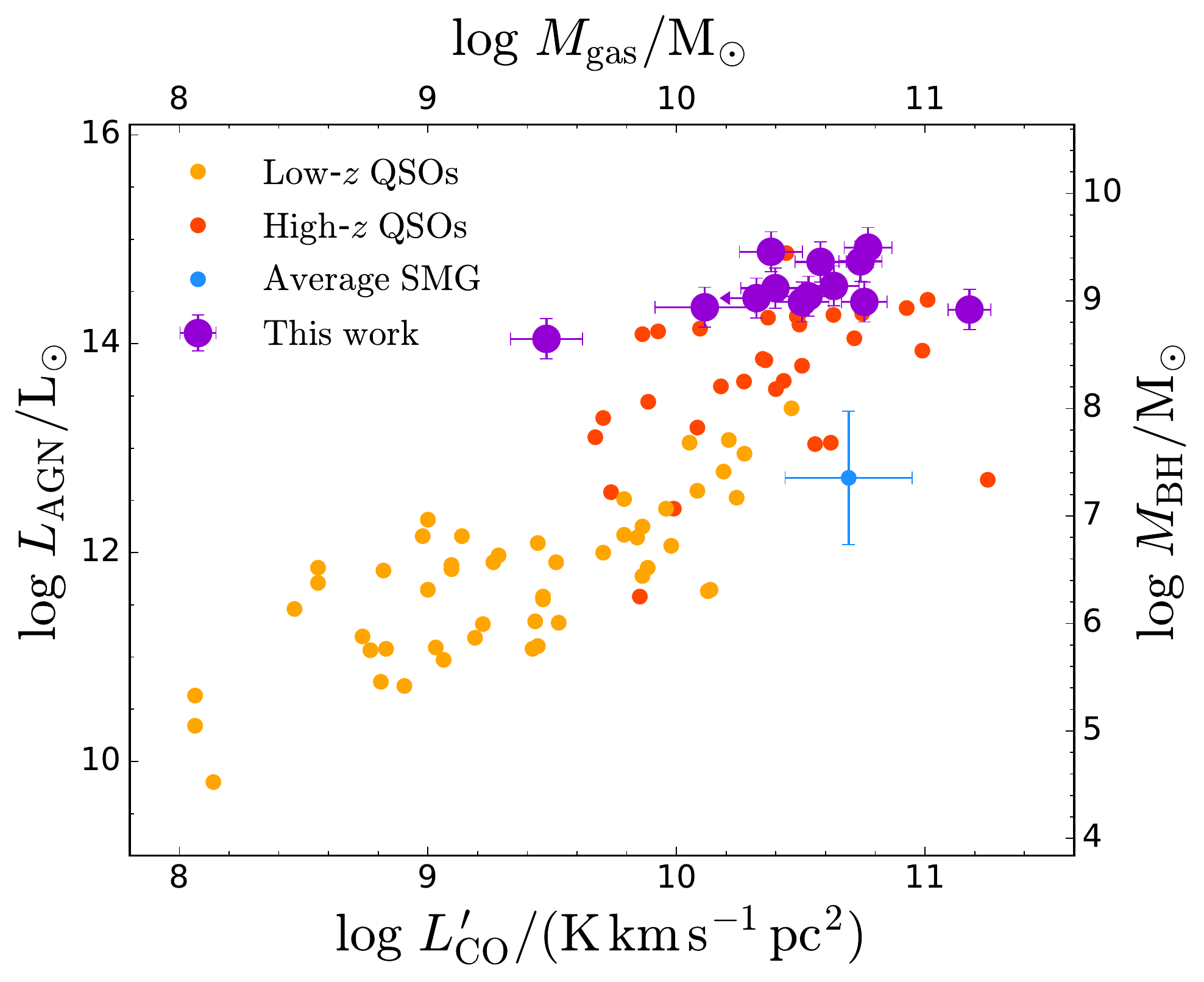}
\caption{AGN luminosity versus line strength for our sample, compared to low-$z$ QSOs and high-$z$ QSOs. Our sources seem to fit the scatter of high-$z$ QSOs well. On the right-hand axis we show the corresponding black hole masses, and on the top axis the gass masses. Shown in blue is the average black-hole mass and gas mass from the SMGs in \citet{bothwell2013}.}
\label{agn}
\end{figure}

As mentioned above, the AGN luminosities provide information about the mass of the central BH within each host galaxy. The BH masses for our sample have been estimated by \citet{trainor2012} by assuming that the 1450\,\AA\ luminosity is completely Eddington-limited so that the BH mass is directly proportional to $L_{1450}$ (with a constant of proportionality of $3.1 \times 10^{-5}$ if the luminosity and mass are in solar units). Now, the AGN luminosity is also proportional to $L_{1450}$ (Eq.~\ref{eq:agn}), which means that we can recast our measurements of $L_{\rm AGN}$ in terms of a BH mass, $M_{\rm BH}$, as follows:
\begin{equation}\label{eq:bh}
M_{\rm BH} = 5.5 \times 10^{-6} \left(L_{\rm AGN} / {\rm L}_{\odot} \right) {\rm M}_{\odot}.
\end{equation}
\noindent
In this equation, the constant of proportionality is the quotient of $3.1 \times 10^{-5}$, the conversion factor of the UV luminosity to BH mass, and 5.6, the conversion factor of UV luminosity to AGN luminosity. In Fig.~\ref{agn} we show the corresponding $M_{\rm BH}$ values on the right-hand axis, and also the corresponding gas mass, $M_{\rm gas}$, on the top axis, obtained from a conversion factor of $L^{\prime}_{\rm CO(1-0)}$ to $M_{\rm gas}$ of 1 in the usual units \citep[e.g.][]{carilli2013}. Lastly, in Table \ref{table:properties}, we provide the values of $L_{\rm AGN}$ and $M_{\rm BH}$ taken directly from \citet{trainor2012}.

\section{Discussion}\label{discussion}
\label{discussion}

One interesting feature of these observations is that we see double peaks in three out of 12 line profiles. Three interpretations are possible, namely mergers, outflows or discs. The first interpretation could help explain the copious submm continuum emission detected in each of these sources, since major mergers are frequently linked to massive bursts of star formation \citep[e.g.,][]{tacconi2008,engel2010,luo2014,chen2015}, which in turn generate large amounts of submm radiation from starlight reprocessed as dust. Therefore, we might expect that merger events in these QSOs are responsible for the 450- and 850-$\mu$m continuum emission detected by SCUBA-2. Moreover, if these are unresolved pairs of galaxies, they may be further systems suitable for tomography. For the second interpretation, molecular outflows have already been observed and studied in local ultra-luminous IR galaxies \citep[ULIRGs;][]{cicone2014}, and this can be used as a mechanism to explain how feedback can quench star formation in galaxies \citep[e.g.,][]{cattaneo2009,fabian2012,dubois2016,pontzen2017}. For the third interpretation, discs are known to produce sharply double-peaked spectra (like those of HS0105$+$1619 and Q2206$-$199) simply due to the differential rotation of gas and the geometry of the system, and they have been observed in rotating discs of some SMGs \citep[e.g.,][]{bothwell2010,hodge2012}. The case of HS1549$+$1919, being composed of two broad peaks, is most likely due to one of the first possibilities (but our data are insufficient to distinguish between them), while those of HS0105$+$1619 and Q2206$-$199, which are composed of two sharp peaks, are perhaps most likely the latter.

In Fig.~\ref{luminosity}, we show $L^{\prime}_{\rm CO(1-0)}/L_{\rm FIR}$, a commonly studied ratio of the parameters that we have derived. This is a proxy for the gas consumption time-scale, since $L^{\prime}_{\rm CO(1-0)}$ is proportional to the gas mass \citep[e.g.][]{bolatto2015} and $L_{\rm FIR}$ is proportional to the star-formation rate \citep[e.g.][]{kennicutt98}. The difference in gas consumption time-scales between QSOs and SMGs was investigated by \citet{simpson2012}, who found that QSOs have shorter time-scales than SMGs by 50$\pm$23\,per cent. In our sample, we find that the ratio of hyper-luminous QSO to SMG mean gas consumption time-scales is 0.45$\pm$0.38, where the uncertainty has been propagated from the standard deviations of the samples, in agreement with this previous result. This difference can be attributed to the observation that the hyper-luminous QSOs have lower $L^{\prime}_{\rm CO(1-0)}$ values at a fixed $L_{\rm FIR}$ by a factor of a few. Nevertheless, the hyper-luminous QSOs in this study do not appear to have very different CO luminosities (or gas masses) compared to typical SMGs. Looking at Fig.~\ref{fwhm}, it is apparent that this is also true for the linewidths (or dynamical masses), where these populations are statistically indistinguishable. Therefore, broadly speaking, the submm properties of these two populations derived from our observations are similar to one another.

Figure \ref{agn} is in sharp contrast with this result if we consider the SMGs from \citet{bothwell2013}, where we also show the average SMG BH mass obtained from measurements of X-ray luminosities. Here we can see that the SMG BH masses are nearly two orders of magnitude lower than those of the hyper-luminous QSOs. Given the extensive evidence for an evolutionary connection between SMGs and QSOs \citep[e.g.,][]{sanders1988,wall2008,coppin2008,simpson2012}, it is difficult to explain the origin of the extremely massive hyper-luminous QSO BHs inferred through rest-frame UV observations. There is additional evidence that BH mass grows in proportion to host galaxy mass \citep[e.g.,][]{kauffmann2000,hopkins2005,dimatteo2005,mullaney2012}, thus we would expect the gas masses or dynamical masses of hyper-luminous QSOs to be proportionally larger than those of SMGs; however, this is not what we see in Figs.~\ref{luminosity} and \ref{fwhm}. 

There are additional observational constraints for these systems from the optical and IR. \citet{trainor2012} found that these hyper-luminous QSOs do not inhabit any more massive haloes than less luminous ($\approx10^{13}\,{\rm L}_{\odot}$) QSOs -- in particular, they found that the fraction of dark matter haloes massive enough to host hyper-luminous QSOs to the number of observed hyper-luminous QSOs is between 10$^{-6}$ and 10$^{-7}$, meaning that they are likely rare events occuring on time-scales of 1--20\,Myr within the larger population of more typical QSOs. \citet{trainor2012} also found that the surrounding regions are over-dense (with $\delta \simeq 7$) and that the galaxy group members have velocity differences of around 200\,km\,s$^{-1}$, making mergers potentially easy to instigate.

There are two obvious ways to explain these observations, namely that the BH masses are correct but selection effects are making these objects appear more incongruous than they really are, or that the BH masses have been overestimated. Let us start with the first explanation. First, we could have selected QSOs that are preferentially face-on by following up only the brightest ones in the rest-frame UV, as has already been seen in optical samples \citep[e.g.][]{carilli2006,alexander2008}. This would mean that the dynamical masses we have derived underestimate the actual masses of the galaxies, making them more compatible with the very large BH masses that come from the UV data. We could have also selected the QSOs that lie on the tail ends of the scatter in various host-galaxy mass-BH mass relationships. During phases of rapid growth, there is evidence that the scatter in the ratio of stellar mass-to-BH mass can be an order of magnitude \citep[McAlpine, priv.~comm.; see also][]{mcalpine2017,mcalpine2018}, so by essentially selecting QSOs with the largest BH masses (via the rest-frame UV luminosity), we could have selected the QSOs with the most extreme host-galaxy mass-to-BH mass ratios. This would additionally explain why these QSOs are seen to have for example more typical gas masses and halo masses.

Now let us examine the possibility that the BH masses have not been calculated properly. Equation \ref{eq:bh} relies on the assumption of Eddington-limited accretion, but this assumption could be wrong. If the QSOs in our sample are undergoing short bursts (about 1--20\,Myr, see \citealt{goncalves2008,trainor2013}) of super-Eddington accretion, this would mean that the actual BH masses are up to a factor of 10 lower than what we have estimated, and hence in agreement with those of SMGs and other high-redshift QSOs  -- then the similar submm properties and halo masses between these populations emerges naturally. Super-Eddington accretion has been claimed in several QSO systems already, particularly in a subclass called narrow-line Seyfert 1 (NLS1) galaxies \citep[e.g.,][]{komossa2006,du2015,lanzuisi2016,jin2017}, while simulations and theoretical calculations are able to reproduce super-Eddington accretion under a sustained infall of cold gas \citep[e.g.,][]{li2012,jiang2017,mayer2018}. As a consistency check, we can estimate the amount by which a BH with a mass of order 10$^8$\,M$_{\odot}$ (in line with what one would expect from average SMGs and QSOs) will grow in 1--20\,Myr assuming an Eddington ratio of $L/L_{\rm Edd}\,{=}\,10$ and a radiative efficiency of order 0.1 \citep[e.g.,][]{davis2011,trakhtenbrot2017}. The total accretion rate in this case would be about 20\,M$_{\odot}$\,yr$^{-1}$, and the total amount of mass accreted over this time-scale would be (0.2--4)$\,{\times}\,$10$^8$\,M$_{\odot}$, which is not unreasonable for final BH masses to remain in the regime of $10^8$\,M$_{\odot}$. 

Given the range of our current data, both of these scenarios appear possible, and the reality might be a mixture of the two. If we wish to better understand these QSOs, more detailed observations would be required; for example, higher resolution imaging may be able to tell us whether or not these QSOs are in fact seen preferentially face-on, or if mergers are present and responsible for some of the double peaks, or if the double peaks are in fact massive molecular outflows. In terms of the overall scope of the SCUBA-2 Web survey, however, this data has certainly provided us with much more detailed information about the far-IR environments of hyper-luminous QSOs.

\section{Summary and conclusion}\label{conclusion}

We have observed 13 of the most UV-luminous QSOs in the submm as part of the SCUBA-2 Web survey using the NOEMA interferometer, targeting CO(3--2) transition lines. These QSOs have also been observed in the submm with SCUBA-2, revealing bright 850-$\mu$m flux densities in every single source. We detected CO(3--2) in 11 of the 13 QSOs; for the remaining two sources, in one we found a tentative 2$\sigma$ line at the expected position and redshift, and in the other we found copious amounts of continuum emission across the band. We also found companion galaxies via CO line emission close to three of the QSOs; these systems are excellent candidates for future tomographic studies of the IGM through measurements of absorption spectra.

We analysed the CO(3--2) line profiles to obtain amplitudes, linewidths, and redshifts. We found that three spectra are well fit by double Gaussians, while the remainder are fit by a single Gaussian. We also calculated line strengths and luminosities. Q2343$+$125 was found to be an extreme object, containing a line strength about 5 times brighter than most of the other sources in our sample, motivating further studies to better understand the circumstances fueling its activity. 

From our line and continuum measurements we derived the quantities $L^{\prime}_{\rm CO(1-0)}$ (which are proportional to gas masses) and $L_{\rm FIR}$. We compared these quantities, along with the linewidths from our fits (which are proportional to dynamical masses), to samples of low redshift QSOs, high redshift QSOs, Type II QSOs and high redshift SMGs from the literature. We found that these submm-derived properties are very similar to other high-redshift QSOs, and also very similar compared to SMGs except for their far-IR luminosities, which are larger by a factor of about 2. This is in contrast with the UV-derived property $L_{\rm AGN}$, where the hyper-luminous QSOs in our sample have some of the largest values ever observed. Since $L_{\rm AGN}$ is proportional to $M_{\rm BH}$ under the assumption of Eddington-limited accretion, we were able to compare BH masses of our QSOs to those of SMGs. We found that our derived BH masses are about two orders of magnitude larger than the average SMG BH mass and one order of magnitude compared to other high-redshift QSOs, at odds with our observations that these QSOs have comparable gas and dynamical masses.

We discuss two interpretations of these data. In the first, the BH masses are correct but selection effects are making these objects appear more anomalous than they really are. The selection we made here was of the most luminous QSOs in the rest-frame UV, and possible selection effects this might induce include preferential face-on viewing angles and picking out objects in the tail ends of the scatter in various host-galaxy mass-BH mass relationships. In the second, the BH masses have been overestimated because the accretion rates are super-Eddington. While the data we have at hand lacks the resolution to distinguish between these scenarios, the reality is probably a combination of the two.

These observations represent just the first phase of the SCUBA-2 Web survey; evidently, the submm and mm observation of these UV-hyper-luminous QSOs have interesting consequences for determining their environments and physical states. More observations of other lines, as well as at higher spatial resolution, will be required to fully understand the nature of these objects, how they relate to their less luminous counterparts, and how they influence their environments, while more observations of the surrounding galaxies should confirm weather or not they can be used for future tomographical studies.

\section*{ACKNOWLEDGMENTS}

The authors would like to thank Christine Done, Stuart McAlpine, and Chris Willott for their helpful discussions. This work is based on observations carried out under project number S17BS with the Institut de Radioastronomie Millim{\'e}trique (IRAM) NOEMA Interferometer. IRAM is supported by the Institut National des Sciences de l'Univers/Centre National de la Recherche Scientifique (INSU/CNRS, France), the Max Planck Gesellschaft (MPG, Germany), and the Instituto Geogr{\'a}fico Nacional (IGN, Spain). The James Clerk Maxwell Telescope is operated by the East Asian Observatory on behalf of The National Astronomical Observatory of Japan; Academia Sinica Institute of Astronomy and Astrophysics; the Korea Astronomy and Space Science Institute; the Operation, Maintenance and Upgrading Fund for Astronomical Telescopes and Facility Instruments, budgeted from the Ministry of Finance (MOF) of China and administrated by the Chinese Academy of Sciences (CAS), as well as the National Key R\&D Program of China (No. 2017YFA0402700). Additional funding support is provided by the Science and Technology Facilities Council of the United Kingdom and participating universities in the United Kingdom and Canada. The authors wish to recognize and acknowledge the very significant cultural role and reverence that the summit of Maunakea has always had within the indigenous Hawaiian community. This work was supported by the Natural Sciences and Engineering Research Council of Canada. Ian Smail acknowledges the European Research Council Advanced Investigator programme DUSTYGAL 321334 and the Science and Technology Facilities Council grant ST/P000541/1. This work is based in part on observations made with the {\it Spitzer Space Telescope}, which is operated by the Jet Propulsion Laboratory, California Institute of Technology under a contract with the National Aeronautics and Space Administration (NASA). Yuichi Matsuda acknowledges the Japan Society for the Promotion of Science (JSPS) KAKENHI grants 17H04831 and 17KK0098. This paper includes data gathered with the 6.5\,m Magellan Telescopes located at Las Campanas Observatory, Chile.

\bibliographystyle{mnras}
\bibliography{kbssQSOco}

\begin{thebibliography}{}
\makeatletter
\relax
\def\mn@urlcharsother{\let\do\@makeother \do\$\do\&\do\#\do\^\do\_\do\%\do\~}
\def\mn@doi{\begingroup\mn@urlcharsother \@ifnextchar [ {\mn@doi@}
  {\mn@doi@[]}}
\def\mn@doi@[#1]#2{\def\@tempa{#1}\ifx\@tempa\@empty \href
  {http://dx.doi.org/#2} {doi:#2}\else \href {http://dx.doi.org/#2} {#1}\fi
  \endgroup}
\def\mn@eprint#1#2{\mn@eprint@#1:#2::\@nil}
\def\mn@eprint@arXiv#1{\href {http://arxiv.org/abs/#1} {{\tt arXiv:#1}}}
\def\mn@eprint@dblp#1{\href {http://dblp.uni-trier.de/rec/bibtex/#1.xml}
  {dblp:#1}}
\def\mn@eprint@#1:#2:#3:#4\@nil{\def\@tempa {#1}\def\@tempb {#2}\def\@tempc
  {#3}\ifx \@tempc \@empty \let \@tempc \@tempb \let \@tempb \@tempa \fi \ifx
  \@tempb \@empty \def\@tempb {arXiv}\fi \@ifundefined
  {mn@eprint@\@tempb}{\@tempb:\@tempc}{\expandafter \expandafter \csname
  mn@eprint@\@tempb\endcsname \expandafter{\@tempc}}}

\bibitem[\protect\citeauthoryear{{Alexander} et~al.,}{{Alexander}
  et~al.}{2008}]{alexander2008}
{Alexander} D.~M.,  et~al., 2008, \mn@doi [\aj] {10.1088/0004-6256/135/5/1968},
  \href {http://adsabs.harvard.edu/abs/2008AJ....135.1968A} {135, 1968}

\bibitem[\protect\citeauthoryear{{Alloin}, {Guilloteau}, {Barvainis},
  {Antonucci}  \& {Tacconi}}{{Alloin} et~al.}{1997}]{alloin1997}
{Alloin} D.,  {Guilloteau} S.,  {Barvainis} R.,  {Antonucci} R.,   {Tacconi}
  L.,  1997, \aap, \href {http://adsabs.harvard.edu/abs/1997A%26A...321...24A}
  {321, 24}

\bibitem[\protect\citeauthoryear{{Anh}, {Boone}, {Hoai}, {Nhung}, {Wei{\ss}},
  {Kneib}, {Beelen}  \& {Salom{\'e}}}{{Anh} et~al.}{2013}]{anh2013}
{Anh} P.~T.,  {Boone} F.,  {Hoai} D.~T.,  {Nhung} P.~T.,  {Wei{\ss}} A.,
  {Kneib} J.~P.,  {Beelen} A.,   {Salom{\'e}} P.,  2013, \mn@doi [\aap]
  {10.1051/0004-6361/201321363}, \href
  {https://ui.adsabs.harvard.edu/\#abs/2013A&A...552L..12A} {552, L12}

\bibitem[\protect\citeauthoryear{{Aravena} et~al.,}{{Aravena}
  et~al.}{2016}]{aravena2016}
{Aravena} M.,  et~al., 2016, \mn@doi [\mnras] {10.1093/mnras/stw275}, \href
  {https://ui.adsabs.harvard.edu/\#abs/2016MNRAS.457.4406A} {457, 4406}

\bibitem[\protect\citeauthoryear{{Asplund}, {Grevesse}, {Sauval}  \&
  {Scott}}{{Asplund} et~al.}{2009}]{Asplund09}
{Asplund} M.,  {Grevesse} N.,  {Sauval} A.~J.,   {Scott} P.,  2009, \mn@doi
  [\araa] {10.1146/annurev.astro.46.060407.145222}, \href
  {http://adsabs.harvard.edu/abs/2009ARA%26A..47..481A} {47, 481}

\bibitem[\protect\citeauthoryear{{Bajtlik}, {Duncan}  \& {Ostriker}}{{Bajtlik}
  et~al.}{1988}]{bajtlik1988}
{Bajtlik} S.,  {Duncan} R.~C.,   {Ostriker} J.~P.,  1988, \mn@doi [\apj]
  {10.1086/166217}, \href {http://adsabs.harvard.edu/abs/1988ApJ...327..570B}
  {327, 570}

\bibitem[\protect\citeauthoryear{{Barkana} \& {Loeb}}{{Barkana} \&
  {Loeb}}{2001}]{barkana2001}
{Barkana} R.,  {Loeb} A.,  2001, \mn@doi [\physrep]
  {10.1016/S0370-1573(01)00019-9}, \href
  {http://adsabs.harvard.edu/abs/2001PhR...349..125B} {349, 125}

\bibitem[\protect\citeauthoryear{{Beelen}, {Cox}, {Benford}, {Dowell},
  {Kov{\'a}cs}, {Bertoldi}, {Omont}  \& {Carilli}}{{Beelen}
  et~al.}{2006}]{beelen2006}
{Beelen} A.,  {Cox} P.,  {Benford} D.~J.,  {Dowell} C.~D.,  {Kov{\'a}cs} A.,
  {Bertoldi} F.,  {Omont} A.,   {Carilli} C.~L.,  2006, \mn@doi [\apj]
  {10.1086/500636}, \href {http://adsabs.harvard.edu/abs/2006ApJ...642..694B}
  {642, 694}

\bibitem[\protect\citeauthoryear{{Berg}, {Neeleman}, {Prochaska}, {Ellison}  \&
  {Wolfe}}{{Berg} et~al.}{2015}]{Berg15}
{Berg} T.~A.~M.,  {Neeleman} M.,  {Prochaska} J.~X.,  {Ellison} S.~L.,
  {Wolfe} A.~M.,  2015, \mn@doi [\pasp] {10.1086/680210}, \href
  {http://adsabs.harvard.edu/abs/2015PASP..127..167B} {127, 167}

\bibitem[\protect\citeauthoryear{{Bertram}, {Eckart}, {Fischer}, {Zuther},
  {Straubmeier}, {Wisotzki}  \& {Krips}}{{Bertram} et~al.}{2007}]{bertram2007}
{Bertram} T.,  {Eckart} A.,  {Fischer} S.,  {Zuther} J.,  {Straubmeier} C.,
  {Wisotzki} L.,   {Krips} M.,  2007, \mn@doi [\aap]
  {10.1051/0004-6361:20077578}, \href
  {http://adsabs.harvard.edu/abs/2007A%26A...470..571B} {470, 571}

\bibitem[\protect\citeauthoryear{{Bolatto}}{{Bolatto}}{2015}]{bolatto2015}
{Bolatto} A.~D.,  2015, in {Simon} R.,  {Schaaf} R.,   {Stutzki} J.,  eds,  EAS
  Publications Series Vol. 75, Conditions and Impact of Star Formation. EDP
  Sciences, Paris, pp 81--86

\bibitem[\protect\citeauthoryear{{Bothwell} et~al.,}{{Bothwell}
  et~al.}{2010}]{bothwell2010}
{Bothwell} M.~S.,  et~al., 2010, \mn@doi [\mnras]
  {10.1111/j.1365-2966.2010.16480.x}, \href
  {http://adsabs.harvard.edu/abs/2010MNRAS.405..219B} {405, 219}

\bibitem[\protect\citeauthoryear{{Bothwell} et~al.,}{{Bothwell}
  et~al.}{2013}]{bothwell2013}
{Bothwell} M.~S.,  et~al., 2013, \mn@doi [\mnras] {10.1093/mnras/sts562}, \href
  {http://adsabs.harvard.edu/abs/2013MNRAS.429.3047B} {429, 3047}

\bibitem[\protect\citeauthoryear{{Bridle} \& {Perley}}{{Bridle} \&
  {Perley}}{1984}]{bridle1984}
{Bridle} A.~H.,  {Perley} R.~A.,  1984, \mn@doi [\araa]
  {10.1146/annurev.aa.22.090184.001535}, \href
  {http://adsabs.harvard.edu/abs/1984ARA%26A..22..319B} {22, 319}

\bibitem[\protect\citeauthoryear{{Carilli} \& {Walter}}{{Carilli} \&
  {Walter}}{2013}]{carilli2013}
{Carilli} C.~L.,  {Walter} F.,  2013, \mn@doi [\araa]
  {10.1146/annurev-astro-082812-140953}, \href
  {http://adsabs.harvard.edu/abs/2013ARA%26A..51..105C} {51, 105}

\bibitem[\protect\citeauthoryear{{Carilli} \& {Wang}}{{Carilli} \&
  {Wang}}{2006}]{carilli2006}
{Carilli} C.~L.,  {Wang} R.,  2006, \mn@doi [\aj] {10.1086/503872}, \href
  {http://adsabs.harvard.edu/abs/2006AJ....131.2763C} {131, 2763}

\bibitem[\protect\citeauthoryear{{Carilli} et~al.,}{{Carilli}
  et~al.}{2002}]{carilli2002}
{Carilli} C.~L.,  et~al., 2002, \mn@doi [\aj] {10.1086/339306}, \href
  {http://adsabs.harvard.edu/abs/2002AJ....123.1838C} {123, 1838}

\bibitem[\protect\citeauthoryear{{Carilli} et~al.,}{{Carilli}
  et~al.}{2007}]{carilli2007}
{Carilli} C.~L.,  et~al., 2007, \mn@doi [\apjl] {10.1086/521648}, \href
  {http://adsabs.harvard.edu/abs/2007ApJ...666L...9C} {666, L9}

\bibitem[\protect\citeauthoryear{{Carniani} et~al.,}{{Carniani}
  et~al.}{2013}]{cariani2013}
{Carniani} S.,  et~al., 2013, \mn@doi [\aap] {10.1051/0004-6361/201322320},
  \href {https://ui.adsabs.harvard.edu/\#abs/2013A&A...559A..29C} {559, A29}

\bibitem[\protect\citeauthoryear{{Cattaneo} et~al.,}{{Cattaneo}
  et~al.}{2009}]{cattaneo2009}
{Cattaneo} A.,  et~al., 2009, \mn@doi [\nat] {10.1038/nature08135}, \href
  {http://adsabs.harvard.edu/abs/2009Natur.460..213C} {460, 213}

\bibitem[\protect\citeauthoryear{{Chen} et~al.,}{{Chen}
  et~al.}{2015}]{chen2015}
{Chen} C.-C.,  et~al., 2015, \mn@doi [\apj] {10.1088/0004-637X/799/2/194},
  \href {http://adsabs.harvard.edu/abs/2015ApJ...799..194C} {799, 194}

\bibitem[\protect\citeauthoryear{{Chenu} et~al.,}{{Chenu}
  et~al.}{2016}]{chenu2016}
{Chenu} J.-Y.,  et~al., 2016, \mn@doi [IEEE Transactions on Terahertz Science
  and Technology] {10.1109/TTHZ.2016.2525762}, \href
  {http://adsabs.harvard.edu/abs/2016ITTST...6..223C} {6, 223}

\bibitem[\protect\citeauthoryear{{Cicone} et~al.,}{{Cicone}
  et~al.}{2014}]{cicone2014}
{Cicone} C.,  et~al., 2014, \mn@doi [\aap] {10.1051/0004-6361/201322464}, \href
  {http://adsabs.harvard.edu/abs/2014A%26A...562A..21C} {562, A21}

\bibitem[\protect\citeauthoryear{{Condon}, {Cotton}, {Greisen}, {Yin},
  {Perley}, {Taylor}  \& {Broderick}}{{Condon} et~al.}{1998}]{condon1998}
{Condon} J.~J.,  {Cotton} W.~D.,  {Greisen} E.~W.,  {Yin} Q.~F.,  {Perley}
  R.~A.,  {Taylor} G.~B.,   {Broderick} J.~J.,  1998, \mn@doi [\aj]
  {10.1086/300337}, \href {http://adsabs.harvard.edu/abs/1998AJ....115.1693C}
  {115, 1693}

\bibitem[\protect\citeauthoryear{{Coppin} et~al.,}{{Coppin}
  et~al.}{2008}]{coppin2008}
{Coppin} K.~E.~K.,  et~al., 2008, \mn@doi [\mnras]
  {10.1111/j.1365-2966.2008.13553.x}, \href
  {http://adsabs.harvard.edu/abs/2008MNRAS.389...45C} {389, 45}

\bibitem[\protect\citeauthoryear{{Dai} et~al.,}{{Dai} et~al.}{2012}]{dai2012}
{Dai} Y.~S.,  et~al., 2012, \mn@doi [\apj] {10.1088/0004-637X/753/1/33}, \href
  {http://adsabs.harvard.edu/abs/2012ApJ...753...33D} {753, 33}

\bibitem[\protect\citeauthoryear{{Dale} \& {Helou}}{{Dale} \&
  {Helou}}{2002}]{dale2002}
{Dale} D.~A.,  {Helou} G.,  2002, \mn@doi [\apj] {10.1086/341632}, \href
  {http://adsabs.harvard.edu/abs/2002ApJ...576..159D} {576, 159}

\bibitem[\protect\citeauthoryear{{Dall'Aglio}, {Wisotzki}  \&
  {Worseck}}{{Dall'Aglio} et~al.}{2008}]{aglio2008}
{Dall'Aglio} A.,  {Wisotzki} L.,   {Worseck} G.,  2008, \mn@doi [\aap]
  {10.1051/0004-6361:20077088}, \href
  {http://adsabs.harvard.edu/abs/2008A%26A...480..359D} {480, 359}

\bibitem[\protect\citeauthoryear{{Davis} \& {Laor}}{{Davis} \&
  {Laor}}{2011}]{davis2011}
{Davis} S.~W.,  {Laor} A.,  2011, \mn@doi [\apj] {10.1088/0004-637X/728/2/98},
  \href {http://adsabs.harvard.edu/abs/2011ApJ...728...98D} {728, 98}

\bibitem[\protect\citeauthoryear{{Di Matteo}, {Springel}  \& {Hernquist}}{{Di
  Matteo} et~al.}{2005}]{dimatteo2005}
{Di Matteo} T.,  {Springel} V.,   {Hernquist} L.,  2005, \mn@doi [\nat]
  {10.1038/nature03335}, \href
  {http://adsabs.harvard.edu/abs/2005Natur.433..604D} {433, 604}

\bibitem[\protect\citeauthoryear{{Du} et~al.,}{{Du} et~al.}{2015}]{du2015}
{Du} P.,  et~al., 2015, \mn@doi [\apj] {10.1088/0004-637X/806/1/22}, \href
  {http://adsabs.harvard.edu/abs/2015ApJ...806...22D} {806, 22}

\bibitem[\protect\citeauthoryear{{Dubois}, {Peirani}, {Pichon}, {Devriendt},
  {Gavazzi}, {Welker}  \& {Volonteri}}{{Dubois} et~al.}{2016}]{dubois2016}
{Dubois} Y.,  {Peirani} S.,  {Pichon} C.,  {Devriendt} J.,  {Gavazzi} R.,
  {Welker} C.,   {Volonteri} M.,  2016, \mn@doi [\mnras]
  {10.1093/mnras/stw2265}, \href
  {http://adsabs.harvard.edu/abs/2016MNRAS.463.3948D} {463, 3948}

\bibitem[\protect\citeauthoryear{{Elbaz} et~al.,}{{Elbaz}
  et~al.}{2010}]{elbaz2010}
{Elbaz} D.,  et~al., 2010, \mn@doi [\aap] {10.1051/0004-6361/201014687}, \href
  {http://adsabs.harvard.edu/abs/2010A%26A...518L..29E} {518, L29}

\bibitem[\protect\citeauthoryear{{Engel} et~al.,}{{Engel}
  et~al.}{2010}]{engel2010}
{Engel} H.,  et~al., 2010, \mn@doi [\apj] {10.1088/0004-637X/724/1/233}, \href
  {http://adsabs.harvard.edu/abs/2010ApJ...724..233E} {724, 233}

\bibitem[\protect\citeauthoryear{{Erb}, {Shapley}, {Steidel}, {Pettini},
  {Adelberger}, {Hunt}, {Moorwood}  \& {Cuby}}{{Erb} et~al.}{2003}]{erb2003}
{Erb} D.~K.,  {Shapley} A.~E.,  {Steidel} C.~C.,  {Pettini} M.,  {Adelberger}
  K.~L.,  {Hunt} M.~P.,  {Moorwood} A.~F.~M.,   {Cuby} J.-G.,  2003, \mn@doi
  [\apj] {10.1086/375316}, \href
  {http://adsabs.harvard.edu/abs/2003ApJ...591..101E} {591, 101}

\bibitem[\protect\citeauthoryear{{Evans}, {Solomon}, {Tacconi}, {Vavilkin}  \&
  {Downes}}{{Evans} et~al.}{2006}]{evans2006}
{Evans} A.~S.,  {Solomon} P.~M.,  {Tacconi} L.~J.,  {Vavilkin} T.,   {Downes}
  D.,  2006, \mn@doi [\aj] {10.1086/508416}, \href
  {http://adsabs.harvard.edu/abs/2006AJ....132.2398E} {132, 2398}

\bibitem[\protect\citeauthoryear{{Fabian}}{{Fabian}}{2012}]{fabian2012}
{Fabian} A.~C.,  2012, \mn@doi [\araa] {10.1146/annurev-astro-081811-125521},
  \href {http://adsabs.harvard.edu/abs/2012ARA%26A..50..455F} {50, 455}

\bibitem[\protect\citeauthoryear{{Fan} et~al.,}{{Fan} et~al.}{2001}]{fan2001}
{Fan} X.,  et~al., 2001, \mn@doi [\aj] {10.1086/318033}, \href
  {http://adsabs.harvard.edu/abs/2001AJ....121...54F} {121, 54}

\bibitem[\protect\citeauthoryear{{Fan}, {Knudsen}, {Fogasy}  \&
  {Drouart}}{{Fan} et~al.}{2018}]{fan2018}
{Fan} L.,  {Knudsen} K.~K.,  {Fogasy} J.,   {Drouart} G.,  2018, \mn@doi
  [\apjl] {10.3847/2041-8213/aab496}, \href
  {http://adsabs.harvard.edu/abs/2018ApJ...856L...5F} {856, L5}

\bibitem[\protect\citeauthoryear{{Feruglio} et~al.,}{{Feruglio}
  et~al.}{2015}]{feruglio2015}
{Feruglio} C.,  et~al., 2015, \mn@doi [\aap] {10.1051/0004-6361/201526020},
  \href {http://adsabs.harvard.edu/abs/2015A%26A...583A..99F} {583, A99}

\bibitem[\protect\citeauthoryear{{Fischer} et~al.,}{{Fischer}
  et~al.}{2010}]{fischer2010}
{Fischer} J.,  et~al., 2010, \mn@doi [\aap] {10.1051/0004-6361/201014676},
  \href {http://adsabs.harvard.edu/abs/2010A%26A...518L..41F} {518, L41}

\bibitem[\protect\citeauthoryear{{Gallerani} et~al.,}{{Gallerani}
  et~al.}{2012}]{gallerani2012}
{Gallerani} S.,  et~al., 2012, \mn@doi [\aap] {10.1051/0004-6361/201118705},
  \href {https://ui.adsabs.harvard.edu/\#abs/2012A&A...543A.114G} {543, A114}

\bibitem[\protect\citeauthoryear{{Gebhardt} et~al.,}{{Gebhardt}
  et~al.}{2000}]{gebhardt2000}
{Gebhardt} K.,  et~al., 2000, \mn@doi [\apjl] {10.1086/312840}, \href
  {http://adsabs.harvard.edu/abs/2000ApJ...539L..13G} {539, L13}

\bibitem[\protect\citeauthoryear{{Gon{\c c}alves}, {Steidel}  \&
  {Pettini}}{{Gon{\c c}alves} et~al.}{2008}]{goncalves2008}
{Gon{\c c}alves} T.~S.,  {Steidel} C.~C.,   {Pettini} M.,  2008, \mn@doi [\apj]
  {10.1086/527313}, \href {http://adsabs.harvard.edu/abs/2008ApJ...676..816G}
  {676, 816}

\bibitem[\protect\citeauthoryear{{Gregg}, {Becker}, {White}, {Helfand},
  {McMahon}  \& {Hook}}{{Gregg} et~al.}{1996}]{gregg1996}
{Gregg} M.~D.,  {Becker} R.~H.,  {White} R.~L.,  {Helfand} D.~J.,  {McMahon}
  R.~G.,   {Hook} I.~M.,  1996, \mn@doi [\aj] {10.1086/118024}, \href
  {http://adsabs.harvard.edu/abs/1996AJ....112..407G} {112, 407}

\bibitem[\protect\citeauthoryear{{Guilloteau} et~al.,}{{Guilloteau}
  et~al.}{1992}]{guilloteau1992}
{Guilloteau} S.,  et~al., 1992, \aap, \href
  {http://adsabs.harvard.edu/abs/1992A%26A...262..624G} {262, 624}

\bibitem[\protect\citeauthoryear{{Hatziminaoglou} et~al.,}{{Hatziminaoglou}
  et~al.}{2010}]{hatziminaoglou2010}
{Hatziminaoglou} E.,  et~al., 2010, \mn@doi [\aap]
  {10.1051/0004-6361/201014679}, \href
  {http://adsabs.harvard.edu/abs/2010A%26A...518L..33H} {518, L33}

\bibitem[\protect\citeauthoryear{{Helou}, {Khan}, {Malek}  \&
  {Boehmer}}{{Helou} et~al.}{1988}]{helou1988}
{Helou} G.,  {Khan} I.~R.,  {Malek} L.,   {Boehmer} L.,  1988, \mn@doi [\apjs]
  {10.1086/191285}, \href {http://adsabs.harvard.edu/abs/1988ApJS...68..151H}
  {68, 151}

\bibitem[\protect\citeauthoryear{{Hodge}, {Carilli}, {Walter}, {de Blok},
  {Riechers}, {Daddi}  \& {Lentati}}{{Hodge} et~al.}{2012}]{hodge2012}
{Hodge} J.~A.,  {Carilli} C.~L.,  {Walter} F.,  {de Blok} W.~J.~G.,  {Riechers}
  D.,  {Daddi} E.,   {Lentati} L.,  2012, \mn@doi [\apj]
  {10.1088/0004-637X/760/1/11}, \href
  {http://adsabs.harvard.edu/abs/2012ApJ...760...11H} {760, 11}

\bibitem[\protect\citeauthoryear{{Holland} et~al.,}{{Holland}
  et~al.}{2013}]{holland2013}
{Holland} W.~S.,  et~al., 2013, \mn@doi [\mnras] {10.1093/mnras/sts612}, \href
  {http://adsabs.harvard.edu/abs/2013MNRAS.430.2513H} {430, 2513}

\bibitem[\protect\citeauthoryear{{Hopkins}, {Hernquist}, {Cox}, {Di Matteo},
  {Martini}, {Robertson}  \& {Springel}}{{Hopkins} et~al.}{2005}]{hopkins2005}
{Hopkins} P.~F.,  {Hernquist} L.,  {Cox} T.~J.,  {Di Matteo} T.,  {Martini} P.,
   {Robertson} B.,   {Springel} V.,  2005, \mn@doi [\apj] {10.1086/432438},
  \href {http://adsabs.harvard.edu/abs/2005ApJ...630..705H} {630, 705}

\bibitem[\protect\citeauthoryear{{Ivison} et~al.,}{{Ivison}
  et~al.}{2007}]{ivison2007}
{Ivison} R.~J.,  et~al., 2007, \mn@doi [\mnras]
  {10.1111/j.1365-2966.2007.12044.x}, \href
  {http://adsabs.harvard.edu/abs/2007MNRAS.380..199I} {380, 199}

\bibitem[\protect\citeauthoryear{{Ivison}, {Papadopoulos}, {Smail}, {Greve},
  {Thomson}, {Xilouris}  \& {Chapman}}{{Ivison} et~al.}{2011}]{ivison2011}
{Ivison} R.~J.,  {Papadopoulos} P.~P.,  {Smail} I.,  {Greve} T.~R.,  {Thomson}
  A.~P.,  {Xilouris} E.~M.,   {Chapman} S.~C.,  2011, \mn@doi [\mnras]
  {10.1111/j.1365-2966.2010.18028.x}, \href
  {http://adsabs.harvard.edu/abs/2011MNRAS.412.1913I} {412, 1913}

\bibitem[\protect\citeauthoryear{{Jiang}, {Stone}  \& {Davis}}{{Jiang}
  et~al.}{2017}]{jiang2017}
{Jiang} Y.-F.,  {Stone} J.,   {Davis} S.~W.,  2017, preprint, \href
  {http://adsabs.harvard.edu/abs/2017arXiv170902845J} {} (\mn@eprint {arXiv}
  {1709.02845})

\bibitem[\protect\citeauthoryear{{Jin}, {Done}  \& {Ward}}{{Jin}
  et~al.}{2017}]{jin2017}
{Jin} C.,  {Done} C.,   {Ward} M.,  2017, \mn@doi [\mnras]
  {10.1093/mnras/stx718}, \href
  {http://adsabs.harvard.edu/abs/2017MNRAS.468.3663J} {468, 3663}

\bibitem[\protect\citeauthoryear{{Kauffmann} \& {Haehnelt}}{{Kauffmann} \&
  {Haehnelt}}{2000}]{kauffmann2000}
{Kauffmann} G.,  {Haehnelt} M.,  2000, \mn@doi [\mnras]
  {10.1046/j.1365-8711.2000.03077.x}, \href
  {http://adsabs.harvard.edu/abs/2000MNRAS.311..576K} {311, 576}

\bibitem[\protect\citeauthoryear{{Kennicutt}}{{Kennicutt}}{1998}]{kennicutt98}
{Kennicutt} Jr. R.~C.,  1998, \mn@doi [\araa] {10.1146/annurev.astro.36.1.189},
  \href {http://cdsads.u-strasbg.fr/abs/1998ARA%26A..36..189K} {36, 189}

\bibitem[\protect\citeauthoryear{{Komossa}, {Voges}, {Xu}, {Mathur}, {Adorf},
  {Lemson}, {Duschl}  \& {Grupe}}{{Komossa} et~al.}{2006}]{komossa2006}
{Komossa} S.,  {Voges} W.,  {Xu} D.,  {Mathur} S.,  {Adorf} H.-M.,  {Lemson}
  G.,  {Duschl} W.~J.,   {Grupe} D.,  2006, \mn@doi [\aj] {10.1086/505043},
  \href {http://adsabs.harvard.edu/abs/2006AJ....132..531K} {132, 531}

\bibitem[\protect\citeauthoryear{{Koprowski}, {Dunlop}, {Micha{\l}owski},
  {Coppin}, {Geach}, {McLure}, {Scott}  \& {van der Werf}}{{Koprowski}
  et~al.}{2017}]{koprowski2017}
{Koprowski} M.~P.,  {Dunlop} J.~S.,  {Micha{\l}owski} M.~J.,  {Coppin}
  K.~E.~K.,  {Geach} J.~E.,  {McLure} R.~J.,  {Scott} D.,   {van der Werf}
  P.~P.,  2017, \mn@doi [\mnras] {10.1093/mnras/stx1843}, \href
  {http://adsabs.harvard.edu/abs/2017MNRAS.471.4155K} {471, 4155}

\bibitem[\protect\citeauthoryear{{Krips}, {Eckart}, {Neri}, {Zuther}, {Downes}
  \& {Scharw{\"a}chter}}{{Krips} et~al.}{2005}]{krips2005}
{Krips} M.,  {Eckart} A.,  {Neri} R.,  {Zuther} J.,  {Downes} D.,
  {Scharw{\"a}chter} J.,  2005, \mn@doi [\aap] {10.1051/0004-6361:20052643},
  \href {https://ui.adsabs.harvard.edu/\#abs/2005A&A...439...75K} {439, 75}

\bibitem[\protect\citeauthoryear{{Krips}, {Neri}  \& {Cox}}{{Krips}
  et~al.}{2012}]{krips2012}
{Krips} M.,  {Neri} R.,   {Cox} P.,  2012, \mn@doi [\apj]
  {10.1088/0004-637X/753/2/135}, \href
  {http://adsabs.harvard.edu/abs/2012ApJ...753..135K} {753, 135}

\bibitem[\protect\citeauthoryear{{Lacaille} et~al.,}{{Lacaille}
  et~al.}{2018}]{lacaille2018}
{Lacaille} K.,  et~al., 2018, preprint, \href
  {http://adsabs.harvard.edu/abs/2018arXiv180906882L} {} (\mn@eprint {arXiv}
  {1809.06882})

\bibitem[\protect\citeauthoryear{{Lanzuisi} et~al.,}{{Lanzuisi}
  et~al.}{2016}]{lanzuisi2016}
{Lanzuisi} G.,  et~al., 2016, \mn@doi [\aap] {10.1051/0004-6361/201628325},
  \href {http://adsabs.harvard.edu/abs/2016A%26A...590A..77L} {590, A77}

\bibitem[\protect\citeauthoryear{{Law}, {Steidel}, {Erb}, {Larkin}, {Pettini},
  {Shapley}  \& {Wright}}{{Law} et~al.}{2009}]{law2009}
{Law} D.~R.,  {Steidel} C.~C.,  {Erb} D.~K.,  {Larkin} J.~E.,  {Pettini} M.,
  {Shapley} A.~E.,   {Wright} S.~A.,  2009, \mn@doi [\apj]
  {10.1088/0004-637X/697/2/2057}, \href
  {https://ui.adsabs.harvard.edu/\#abs/2009ApJ...697.2057L} {697, 2057}

\bibitem[\protect\citeauthoryear{{Leipski} et~al.,}{{Leipski}
  et~al.}{2014}]{leipski2014}
{Leipski} C.,  et~al., 2014, \mn@doi [\apj] {10.1088/0004-637X/785/2/154},
  \href {http://adsabs.harvard.edu/abs/2014ApJ...785..154L} {785, 154}

\bibitem[\protect\citeauthoryear{{Leung}, {Riechers}  \& {Pavesi}}{{Leung}
  et~al.}{2017}]{leung2017}
{Leung} T.~K.~D.,  {Riechers} D.~A.,   {Pavesi} R.,  2017, \mn@doi [\apj]
  {10.3847/1538-4357/aa5b98}, \href
  {https://ui.adsabs.harvard.edu/\#abs/2017ApJ...836..180L} {836, 180}

\bibitem[\protect\citeauthoryear{{Li}}{{Li}}{2012}]{li2012}
{Li} L.-X.,  2012, \mn@doi [\mnras] {10.1111/j.1365-2966.2012.21336.x}, \href
  {http://adsabs.harvard.edu/abs/2012MNRAS.424.1461L} {424, 1461}

\bibitem[\protect\citeauthoryear{{Liszt}, {Pety}  \& {Lucas}}{{Liszt}
  et~al.}{2010}]{liszt2010}
{Liszt} H.~S.,  {Pety} J.,   {Lucas} R.,  2010, \mn@doi [\aap]
  {10.1051/0004-6361/201014510}, \href
  {http://adsabs.harvard.edu/abs/2010A%26A...518A..45L} {518, A45}

\bibitem[\protect\citeauthoryear{{Luo}, {Yang}  \& {Zhang}}{{Luo}
  et~al.}{2014}]{luo2014}
{Luo} W.,  {Yang} X.,   {Zhang} Y.,  2014, \mn@doi [\apjl]
  {10.1088/2041-8205/789/1/L16}, \href
  {http://adsabs.harvard.edu/abs/2014ApJ...789L..16L} {789, L16}

\bibitem[\protect\citeauthoryear{{Magorrian} et~al.,}{{Magorrian}
  et~al.}{1998}]{magorrian1998}
{Magorrian} J.,  et~al., 1998, \mn@doi [\aj] {10.1086/300353}, \href
  {http://adsabs.harvard.edu/abs/1998AJ....115.2285M} {115, 2285}

\bibitem[\protect\citeauthoryear{{Maiolino} et~al.,}{{Maiolino}
  et~al.}{2007}]{maiolino2007}
{Maiolino} R.,  et~al., 2007, \mn@doi [\aap] {10.1051/0004-6361:20078136},
  \href {http://adsabs.harvard.edu/abs/2007A%26A...472L..33M} {472, L33}

\bibitem[\protect\citeauthoryear{{Marsden}, {Borys}, {Chapman}, {Halpern}  \&
  {Scott}}{{Marsden} et~al.}{2005}]{marsden2005}
{Marsden} G.,  {Borys} C.,  {Chapman} S.~C.,  {Halpern} M.,   {Scott} D.,
  2005, \mn@doi [\mnras] {10.1111/j.1365-2966.2005.08837.x}, \href
  {http://adsabs.harvard.edu/abs/2005MNRAS.359...43M} {359, 43}

\bibitem[\protect\citeauthoryear{{Mayer}}{{Mayer}}{2018}]{mayer2018}
{Mayer} L.,  2018, preprint, \href
  {http://adsabs.harvard.edu/abs/2018arXiv180706243M} {} (\mn@eprint {arXiv}
  {1807.06243})

\bibitem[\protect\citeauthoryear{{McAlpine}, {Bower}, {Harrison}, {Crain},
  {Schaller}, {Schaye}  \& {Theuns}}{{McAlpine} et~al.}{2017}]{mcalpine2017}
{McAlpine} S.,  {Bower} R.~G.,  {Harrison} C.~M.,  {Crain} R.~A.,  {Schaller}
  M.,  {Schaye} J.,   {Theuns} T.,  2017, \mn@doi [\mnras]
  {10.1093/mnras/stx658}, \href
  {http://adsabs.harvard.edu/abs/2017MNRAS.468.3395M} {468, 3395}

\bibitem[\protect\citeauthoryear{{McAlpine}, {Bower}, {Rosario}, {Crain},
  {Schaye}  \& {Theuns}}{{McAlpine} et~al.}{2018}]{mcalpine2018}
{McAlpine} S.,  {Bower} R.~G.,  {Rosario} D.~J.,  {Crain} R.~A.,  {Schaye} J.,
   {Theuns} T.,  2018, \mn@doi [\mnras] {10.1093/mnras/sty2489}, \href
  {http://adsabs.harvard.edu/abs/2018MNRAS.481.3118M} {481, 3118}

\bibitem[\protect\citeauthoryear{{Micha{\l}owski}, {Murphy}, {Hjorth},
  {Watson}, {Gall}  \& {Dunlop}}{{Micha{\l}owski}
  et~al.}{2010}]{michalowski2010}
{Micha{\l}owski} M.~J.,  {Murphy} E.~J.,  {Hjorth} J.,  {Watson} D.,  {Gall}
  C.,   {Dunlop} J.~S.,  2010, \mn@doi [\aap] {10.1051/0004-6361/201014902},
  \href {https://ui.adsabs.harvard.edu/\#abs/2010A&A...522A..15M} {522, A15}

\bibitem[\protect\citeauthoryear{{Mullaney} et~al.,}{{Mullaney}
  et~al.}{2012}]{mullaney2012}
{Mullaney} J.~R.,  et~al., 2012, \mn@doi [\apjl] {10.1088/2041-8205/753/2/L30},
  \href {http://adsabs.harvard.edu/abs/2012ApJ...753L..30M} {753, L30}

\bibitem[\protect\citeauthoryear{{Papadopoulos}, {van der Werf}, {Xilouris},
  {Isaak}  \& {Gao}}{{Papadopoulos} et~al.}{2012}]{papadopoulos2012}
{Papadopoulos} P.~P.,  {van der Werf} P.,  {Xilouris} E.,  {Isaak} K.~G.,
  {Gao} Y.,  2012, \mn@doi [\apj] {10.1088/0004-637X/751/1/10}, \href
  {http://adsabs.harvard.edu/abs/2012ApJ...751...10P} {751, 10}

\bibitem[\protect\citeauthoryear{{Perrotta} et~al.,}{{Perrotta}
  et~al.}{2016}]{perrotta2016}
{Perrotta} S.,  et~al., 2016, \mn@doi [\mnras] {10.1093/mnras/stw1703}, \href
  {http://adsabs.harvard.edu/abs/2016MNRAS.462.3285P} {462, 3285}

\bibitem[\protect\citeauthoryear{{Planck Collaboration XIII}}{{Planck
  Collaboration XIII}}{2016}]{planck2014-a15}
{Planck Collaboration XIII} 2016, \mn@doi [\aap] {10.1051/0004-6361/201525830},
  594, A13

\bibitem[\protect\citeauthoryear{{Pontzen}, {Tremmel}, {Roth}, {Peiris},
  {Saintonge}, {Volonteri}, {Quinn}  \& {Governato}}{{Pontzen}
  et~al.}{2017}]{pontzen2017}
{Pontzen} A.,  {Tremmel} M.,  {Roth} N.,  {Peiris} H.~V.,  {Saintonge} A.,
  {Volonteri} M.,  {Quinn} T.,   {Governato} F.,  2017, \mn@doi [\mnras]
  {10.1093/mnras/stw2627}, \href
  {http://adsabs.harvard.edu/abs/2017MNRAS.465..547P} {465, 547}

\bibitem[\protect\citeauthoryear{{Price} et~al.,}{{Price}
  et~al.}{2016}]{price2016}
{Price} S.~H.,  et~al., 2016, \mn@doi [\apj] {10.3847/0004-637X/819/1/80},
  \href {https://ui.adsabs.harvard.edu/\#abs/2016ApJ...819...80P} {819, 80}

\bibitem[\protect\citeauthoryear{{Richards}, {Vanden Berk}, {Reichard}, {Hall},
  {Schneider}, {SubbaRao}, {Thakar}  \& {York}}{{Richards}
  et~al.}{2002}]{richards2002}
{Richards} G.~T.,  {Vanden Berk} D.~E.,  {Reichard} T.~A.,  {Hall} P.~B.,
  {Schneider} D.~P.,  {SubbaRao} M.,  {Thakar} A.~R.,   {York} D.~G.,  2002,
  \mn@doi [\aj] {10.1086/341167}, \href
  {http://adsabs.harvard.edu/abs/2002AJ....124....1R} {124, 1}

\bibitem[\protect\citeauthoryear{{Riechers}, {Walter}, {Carilli}, {Bertoldi}
  \& {Momjian}}{{Riechers} et~al.}{2008}]{riechers2008}
{Riechers} D.~A.,  {Walter} F.,  {Carilli} C.~L.,  {Bertoldi} F.,   {Momjian}
  E.,  2008, \mn@doi [\apj] {10.1086/592834}, \href
  {https://ui.adsabs.harvard.edu/\#abs/2008ApJ...686L...9R} {686, L9}

\bibitem[\protect\citeauthoryear{{Rodr{\'{\i}}guez}, {Villar-Mart{\'{\i}}n},
  {Emonts}, {Humphrey}, {Drouart}, {Garc{\'{\i}}a Burillo}  \& {P{\'e}rez
  Torres}}{{Rodr{\'{\i}}guez} et~al.}{2014}]{rodriguez2014}
{Rodr{\'{\i}}guez} M.~I.,  {Villar-Mart{\'{\i}}n} M.,  {Emonts} B.,  {Humphrey}
  A.,  {Drouart} G.,  {Garc{\'{\i}}a Burillo} S.,   {P{\'e}rez Torres} M.,
  2014, \mn@doi [\aap] {10.1051/0004-6361/201323004}, \href
  {http://adsabs.harvard.edu/abs/2014A%26A...565A..19R} {565, A19}

\bibitem[\protect\citeauthoryear{{Rudie} et~al.,}{{Rudie}
  et~al.}{2012}]{rudie2012}
{Rudie} G.~C.,  et~al., 2012, \mn@doi [\apj] {10.1088/0004-637X/750/1/67},
  \href {http://adsabs.harvard.edu/abs/2012ApJ...750...67R} {750, 67}

\bibitem[\protect\citeauthoryear{{Salom{\'e}}, {Gu{\'e}lin}, {Downes}, {Cox},
  {Guilloteau}, {Omont}, {Gavazzi}  \& {Neri}}{{Salom{\'e}}
  et~al.}{2012}]{salome2012}
{Salom{\'e}} P.,  {Gu{\'e}lin} M.,  {Downes} D.,  {Cox} P.,  {Guilloteau} S.,
  {Omont} A.,  {Gavazzi} R.,   {Neri} R.,  2012, \mn@doi [\aap]
  {10.1051/0004-6361/201219955}, \href
  {https://ui.adsabs.harvard.edu/\#abs/2012A&A...545A..57S} {545, A57}

\bibitem[\protect\citeauthoryear{{Sanders}, {Soifer}, {Elias}, {Neugebauer}  \&
  {Matthews}}{{Sanders} et~al.}{1988}]{sanders1988}
{Sanders} D.~B.,  {Soifer} B.~T.,  {Elias} J.~H.,  {Neugebauer} G.,
  {Matthews} K.,  1988, \mn@doi [\apjl] {10.1086/185155}, \href
  {http://adsabs.harvard.edu/abs/1988ApJ...328L..35S} {328, L35}

\bibitem[\protect\citeauthoryear{{Santini} et~al.,}{{Santini}
  et~al.}{2010}]{santini2010}
{Santini} P.,  et~al., 2010, \mn@doi [\aap] {10.1051/0004-6361/201014748},
  \href {https://ui.adsabs.harvard.edu/\#abs/2010A&A...518L.154S} {518, L154}

\bibitem[\protect\citeauthoryear{{Savage} \& {Sembach}}{{Savage} \&
  {Sembach}}{1991}]{Savage91}
{Savage} B.~D.,  {Sembach} K.~R.,  1991, \mn@doi [\apj] {10.1086/170498}, \href
  {http://adsabs.harvard.edu/abs/1991ApJ...379..245S} {379, 245}

\bibitem[\protect\citeauthoryear{{Schmidt}, {Schneider}  \& {Gunn}}{{Schmidt}
  et~al.}{1995}]{schmidt1995}
{Schmidt} M.,  {Schneider} D.~P.,   {Gunn} J.~E.,  1995, \mn@doi [\aj]
  {10.1086/117497}, \href {http://adsabs.harvard.edu/abs/1995AJ....110...68S}
  {110, 68}

\bibitem[\protect\citeauthoryear{{Schumacher}, {Mart{\'\i}nez-Sansigre},
  {Lacy}, {Rawlings}  \& {Schinnerer}}{{Schumacher}
  et~al.}{2012}]{schumacher2012}
{Schumacher} H.,  {Mart{\'\i}nez-Sansigre} A.,  {Lacy} M.,  {Rawlings} S.,
  {Schinnerer} E.,  2012, \mn@doi [\mnras] {10.1111/j.1365-2966.2012.21024.x},
  \href {https://ui.adsabs.harvard.edu/\#abs/2012MNRAS.423.2132S} {423, 2132}

\bibitem[\protect\citeauthoryear{{Scott}, {Bechtold}, {Dobrzycki}  \&
  {Kulkarni}}{{Scott} et~al.}{2000}]{scott2000}
{Scott} J.,  {Bechtold} J.,  {Dobrzycki} A.,   {Kulkarni} V.~P.,  2000, \mn@doi
  [\apjs] {10.1086/317340}, \href
  {http://adsabs.harvard.edu/abs/2000ApJS..130...67S} {130, 67}

\bibitem[\protect\citeauthoryear{{Scoville}, {Frayer}, {Schinnerer}  \&
  {Christopher}}{{Scoville} et~al.}{2003}]{scoville2003}
{Scoville} N.~Z.,  {Frayer} D.~T.,  {Schinnerer} E.,   {Christopher} M.,  2003,
  \mn@doi [\apjl] {10.1086/374544}, \href
  {http://adsabs.harvard.edu/abs/2003ApJ...585L.105S} {585, L105}

\bibitem[\protect\citeauthoryear{{Simpson} et~al.,}{{Simpson}
  et~al.}{2012}]{simpson2012}
{Simpson} J.~M.,  et~al., 2012, \mn@doi [\mnras]
  {10.1111/j.1365-2966.2012.21941.x}, \href
  {http://adsabs.harvard.edu/abs/2012MNRAS.426.3201S} {426, 3201}

\bibitem[\protect\citeauthoryear{{Sluse}, {Hutsem{\'e}kers}, {Courbin},
  {Meylan}  \& {Wambsganss}}{{Sluse} et~al.}{2012}]{sluse2012}
{Sluse} D.,  {Hutsem{\'e}kers} D.,  {Courbin} F.,  {Meylan} G.,   {Wambsganss}
  J.,  2012, \mn@doi [\aap] {10.1051/0004-6361/201219125}, \href
  {http://adsabs.harvard.edu/abs/2012A%26A...544A..62S} {544, A62}

\bibitem[\protect\citeauthoryear{{Sobral}, {Smail}, {Best}, {Geach}, {Matsuda},
  {Stott}, {Cirasuolo}  \& {Kurk}}{{Sobral} et~al.}{2013}]{sobral2013}
{Sobral} D.,  {Smail} I.,  {Best} P.~N.,  {Geach} J.~E.,  {Matsuda} Y.,
  {Stott} J.~P.,  {Cirasuolo} M.,   {Kurk} J.,  2013, \mn@doi [\mnras]
  {10.1093/mnras/sts096}, \href
  {http://adsabs.harvard.edu/abs/2013MNRAS.428.1128S} {428, 1128}

\bibitem[\protect\citeauthoryear{{Solomon} \& {Barrett}}{{Solomon} \&
  {Barrett}}{1991}]{solomon1991}
{Solomon} P.~M.,  {Barrett} J.~W.,  1991, in {Combes} F.,  {Casoli} F.,  eds,
  IAU Symposium Vol. 146, Dynamics of Galaxies and Their Molecular Cloud
  Distributions. Kluwer Academic Publishers, Dordrecht, p.~235

\bibitem[\protect\citeauthoryear{{Solomon} \& {Vanden Bout}}{{Solomon} \&
  {Vanden Bout}}{2005}]{solomon2005}
{Solomon} P.~M.,  {Vanden Bout} P.~A.,  2005, \mn@doi [\araa]
  {10.1146/annurev.astro.43.051804.102221}, \href
  {http://adsabs.harvard.edu/abs/2005ARA%26A..43..677S} {43, 677}

\bibitem[\protect\citeauthoryear{{Solomon}, {Downes}, {Radford}  \&
  {Barrett}}{{Solomon} et~al.}{1997}]{solomon1997}
{Solomon} P.~M.,  {Downes} D.,  {Radford} S.~J.~E.,   {Barrett} J.~W.,  1997,
  \mn@doi [\apj] {10.1086/303765}, \href
  {http://adsabs.harvard.edu/abs/1997ApJ...478..144S} {478, 144}

\bibitem[\protect\citeauthoryear{{Spilker} et~al.,}{{Spilker}
  et~al.}{2018}]{spilker2018}
{Spilker} J.~S.,  et~al., 2018, \mn@doi [Science] {10.1126/science.aap8900},
  \href {http://adsabs.harvard.edu/abs/2018Sci...361.1016S} {361, 1016}

\bibitem[\protect\citeauthoryear{{Spoon} et~al.,}{{Spoon}
  et~al.}{2013}]{spoon2013}
{Spoon} H.~W.~W.,  et~al., 2013, \mn@doi [\apj] {10.1088/0004-637X/775/2/127},
  \href {http://adsabs.harvard.edu/abs/2013ApJ...775..127S} {775, 127}

\bibitem[\protect\citeauthoryear{{Steidel} et~al.,}{{Steidel}
  et~al.}{2014}]{steidel2014}
{Steidel} C.~C.,  et~al., 2014, \mn@doi [\apj] {10.1088/0004-637X/795/2/165},
  \href {http://adsabs.harvard.edu/abs/2014ApJ...795..165S} {795, 165}

\bibitem[\protect\citeauthoryear{{Sun}, {Greene}, {Zakamska}  \&
  {Nesvadba}}{{Sun} et~al.}{2014}]{sun2014}
{Sun} A.-L.,  {Greene} J.~E.,  {Zakamska} N.~L.,   {Nesvadba} N.~P.~H.,  2014,
  \mn@doi [\apj] {10.1088/0004-637X/790/2/160}, \href
  {http://adsabs.harvard.edu/abs/2014ApJ...790..160S} {790, 160}

\bibitem[\protect\citeauthoryear{{Swinbank} et~al.,}{{Swinbank}
  et~al.}{2014}]{swinbank2014}
{Swinbank} A.~M.,  et~al., 2014, \mn@doi [\mnras] {10.1093/mnras/stt2273},
  \href {http://adsabs.harvard.edu/abs/2014MNRAS.438.1267S} {438, 1267}

\bibitem[\protect\citeauthoryear{{Tacconi} et~al.,}{{Tacconi}
  et~al.}{2006}]{tacconi2006}
{Tacconi} L.~J.,  et~al., 2006, \mn@doi [\apj] {10.1086/499933}, \href
  {https://ui.adsabs.harvard.edu/\#abs/2006ApJ...640..228T} {640, 228}

\bibitem[\protect\citeauthoryear{{Tacconi} et~al.,}{{Tacconi}
  et~al.}{2008}]{tacconi2008}
{Tacconi} L.~J.,  et~al., 2008, \mn@doi [\apj] {10.1086/587168}, \href
  {http://adsabs.harvard.edu/abs/2008ApJ...680..246T} {680, 246}

\bibitem[\protect\citeauthoryear{{Tacconi} et~al.,}{{Tacconi}
  et~al.}{2013}]{tacconi2013}
{Tacconi} L.~J.,  et~al., 2013, \mn@doi [\apj] {10.1088/0004-637X/768/1/74},
  \href {https://ui.adsabs.harvard.edu/\#abs/2013ApJ...768...74T} {768, 74}

\bibitem[\protect\citeauthoryear{{Trainor} \& {Steidel}}{{Trainor} \&
  {Steidel}}{2012}]{trainor2012}
{Trainor} R.~F.,  {Steidel} C.~C.,  2012, \mn@doi [\apj]
  {10.1088/0004-637X/752/1/39}, \href
  {http://adsabs.harvard.edu/abs/2012ApJ...752...39T} {752, 39}

\bibitem[\protect\citeauthoryear{{Trainor} \& {Steidel}}{{Trainor} \&
  {Steidel}}{2013}]{trainor2013}
{Trainor} R.,  {Steidel} C.~C.,  2013, \mn@doi [\apjl]
  {10.1088/2041-8205/775/1/L3}, \href
  {http://adsabs.harvard.edu/abs/2013ApJ...775L...3T} {775, L3}

\bibitem[\protect\citeauthoryear{{Trakhtenbrot}, {Volonteri}  \&
  {Natarajan}}{{Trakhtenbrot} et~al.}{2017}]{trakhtenbrot2017}
{Trakhtenbrot} B.,  {Volonteri} M.,   {Natarajan} P.,  2017, \mn@doi [\apjl]
  {10.3847/2041-8213/836/1/L1}, \href
  {http://adsabs.harvard.edu/abs/2017ApJ...836L...1T} {836, L1}

\bibitem[\protect\citeauthoryear{{Veilleux}, {Cecil}  \&
  {Bland-Hawthorn}}{{Veilleux} et~al.}{2005}]{veilleaux2005}
{Veilleux} S.,  {Cecil} G.,   {Bland-Hawthorn} J.,  2005, \mn@doi [\araa]
  {10.1146/annurev.astro.43.072103.150610}, \href
  {http://adsabs.harvard.edu/abs/2005ARA%26A..43..769V} {43, 769}

\bibitem[\protect\citeauthoryear{{Veilleux} et~al.,}{{Veilleux}
  et~al.}{2013}]{veilleux2013}
{Veilleux} S.,  et~al., 2013, \mn@doi [\apj] {10.1088/0004-637X/776/1/27},
  \href {http://adsabs.harvard.edu/abs/2013ApJ...776...27V} {776, 27}

\bibitem[\protect\citeauthoryear{{Verdier}, {Melin}, {Bartlett}, {Magneville},
  {Palanque-Delabrouille}  \& {Y{\`e}che}}{{Verdier}
  et~al.}{2016}]{verdier2016}
{Verdier} L.,  {Melin} J.-B.,  {Bartlett} J.~G.,  {Magneville} C.,
  {Palanque-Delabrouille} N.,   {Y{\`e}che} C.,  2016, \mn@doi [\aap]
  {10.1051/0004-6361/201527431}, \href
  {http://adsabs.harvard.edu/abs/2016A%26A...588A..61V} {588, A61}

\bibitem[\protect\citeauthoryear{{Vestergaard}}{{Vestergaard}}{2004}]{vestergaard2004}
{Vestergaard} M.,  2004, \mn@doi [\apj] {10.1086/379758}, \href
  {http://adsabs.harvard.edu/abs/2004ApJ...601..676V} {601, 676}

\bibitem[\protect\citeauthoryear{{Villar-Mart{\'{\i}}n}
  et~al.,}{{Villar-Mart{\'{\i}}n} et~al.}{2013}]{villar-martin2013}
{Villar-Mart{\'{\i}}n} M.,  et~al., 2013, \mn@doi [\mnras]
  {10.1093/mnras/stt1014}, \href
  {http://adsabs.harvard.edu/abs/2013MNRAS.434..978V} {434, 978}

\bibitem[\protect\citeauthoryear{{Wall}, {Pope}  \& {Scott}}{{Wall}
  et~al.}{2008}]{wall2008}
{Wall} J.~V.,  {Pope} A.,   {Scott} D.,  2008, \mn@doi [\mnras]
  {10.1111/j.1365-2966.2007.12547.x}, \href
  {http://adsabs.harvard.edu/abs/2008MNRAS.383..435W} {383, 435}

\bibitem[\protect\citeauthoryear{{Walter} et~al.,}{{Walter}
  et~al.}{2003}]{walter2003}
{Walter} F.,  et~al., 2003, \mn@doi [\nat] {10.1038/nature01821}, \href
  {http://adsabs.harvard.edu/abs/2003Natur.424..406W} {424, 406}

\bibitem[\protect\citeauthoryear{{Wang} et~al.,}{{Wang}
  et~al.}{2008}]{wang2008}
{Wang} R.,  et~al., 2008, \mn@doi [\aj] {10.1088/0004-6256/135/4/1201}, \href
  {http://adsabs.harvard.edu/abs/2008AJ....135.1201W} {135, 1201}

\bibitem[\protect\citeauthoryear{{Wang} et~al.,}{{Wang}
  et~al.}{2010}]{wang2010}
{Wang} R.,  et~al., 2010, \mn@doi [\apj] {10.1088/0004-637X/714/1/699}, \href
  {http://adsabs.harvard.edu/abs/2010ApJ...714..699W} {714, 699}

\bibitem[\protect\citeauthoryear{{Wang} et~al.,}{{Wang}
  et~al.}{2016}]{wang2016}
{Wang} R.,  et~al., 2016, \mn@doi [\apj] {10.3847/0004-637X/830/1/53}, \href
  {https://ui.adsabs.harvard.edu/\#abs/2016ApJ...830...53W} {830, 53}

\bibitem[\protect\citeauthoryear{{Wiklind} et~al.,}{{Wiklind}
  et~al.}{2014}]{wiklind2014}
{Wiklind} T.,  et~al., 2014, \mn@doi [\apj] {10.1088/0004-637X/785/2/111},
  \href {https://ui.adsabs.harvard.edu/\#abs/2014ApJ...785..111W} {785, 111}

\bibitem[\protect\citeauthoryear{{Xia} et~al.,}{{Xia} et~al.}{2012}]{xia2012}
{Xia} X.~Y.,  et~al., 2012, \mn@doi [\apj] {10.1088/0004-637X/750/2/92}, \href
  {http://adsabs.harvard.edu/abs/2012ApJ...750...92X} {750, 92}

\bibitem[\protect\citeauthoryear{{Yun}, {Reddy}  \& {Condon}}{{Yun}
  et~al.}{2001}]{yun2001}
{Yun} M.~S.,  {Reddy} N.~A.,   {Condon} J.~J.,  2001, \mn@doi [\apj]
  {10.1086/323145}, \href {http://adsabs.harvard.edu/abs/2001ApJ...554..803Y}
  {554, 803}

\bibitem[\protect\citeauthoryear{{Zensus}}{{Zensus}}{1997}]{zensus1997}
{Zensus} J.~A.,  1997, \mn@doi [\araa] {10.1146/annurev.astro.35.1.607}, \href
  {http://adsabs.harvard.edu/abs/1997ARA%26A..35..607Z} {35, 607}

\makeatother
\end{thebibliography}

\appendix

\section{Comments on individual fields}
\label{appendix}
Below we provide more details for each field based on existing KBSS data \citep[e.g.,][]{rudie2012,trainor2012}. We also summarize preliminary results from a search along four QSO sightlines (HS0105$+$1619, Q0142$-$100, HS1700$+$64, and Q2343$+$125) for absorption lines, including their H\textsc{i} column densities, log\,N(H\textsc{i}) (as determined by Voigt profile fitting the Ly$\alpha$ and Ly$\beta$ absorption lines simultaneously) and metallicities \citep[following][]{Savage91} on the \cite{Asplund09} solar scale.
\newline
\linebreak
\textbf{Q0100$+$130} -- 
There is a known continuum-selected galaxy at the same redshift as the QSO about 5\,arcsec to the south, which is not seen in our observations, as this companion galaxy has a significantly lower far-IR luminosity than Q0100$+$130 (suggesting that it would also have relatively weak CO emission). Mapping in rest-frame Ly\,$\alpha$ showed six objects within about 30\,arcsec of Q0100$+$130, as well as extended, lumpy emission from the QSO itself.
\newline
\linebreak
\textbf{HS0105$+$1619} -- 
In this field we detected a companion galaxy in the continuum, which we note is also seen in the NRAO VLA Sky Survey \citep[NVSS;][]{condon1998} radio continuum source catalogue (18.2$\pm$0.7\,mJy at 1.4\,GHz). Since this companion lacks a CO(3--2) line but is bright in the radio, it is likely a field blazar at a different redshift. Two additional galaxies are known to be at the same redshift as HS0105$+$1619 and located within a few arcseconds, however these are not detected at 3\,mm.

There are two intervening Ly$\alpha$ absorbing systems seen towards HS0105$+$1619, with H\textsc{i} column densities of log\,N(H\textsc{i})$=18.1\pm0.2$ ($z=2.1498$) and log\,N(H\textsc{i}) $=19.4\pm0.2$ ($z=2.3137$). The C\textsc{iv} doublet at 1548\,\AA\ and 1550\,\AA\ are detected in both systems (log\,N(C\textsc{iv}) $=12.97\pm0.05$ and $13.22\pm0.05$, respectively). The sub-damped Ly$\alpha$ absorber at $z=2.3137$ has a metallicity of [O/H]$=-1.95\pm0.25$ (based on the O\textsc{i} 1302\,\AA\ line).
\newline
\linebreak
\textbf{Q0142$-$100} -- This source is known to be a doubly-imaged gravitational lens with a separation of 2.2\,arcsec. One image is magnified by a factor of 3.2, and the other image by a factor of 0.4 \citep{sluse2012}. Our NOEMA detection is centred on the more luminous of the two images, and the synthesized beam is small enough that the fainter image does not contribute any flux density. Since the magnification factor for this image is 3.2, Q0142$−$10 is not intrinsically as luminous as it appears. The majority of the mm emission seen in our observation is not being emitted by the central QSO but instead by a nearby companion galaxy, whose spectrum contains two broad peaks. In Fig.~\ref{fig:maps_zoom} we can see that the companion galaxy's two peaks lie between the QSO and another source to the south that is bright in the IR; however, this southern source is at a much lower redshift (about 0.1--0.2) and so not associated with the companion galaxy. 

A damped Ly$\alpha$ system is detected along the QSO sightline of Q0142$-$100, with a column density of log\,N(H\textsc{i}) $=20.6\pm0.2$ ($z=1.6258$). Due to its low redshift, we cannot identify any prominent metal lines as they are all blended in the Ly$\alpha$ forest. Therefore, we estimate the metallicity of the system to be [Cr/H] $=-1.5\pm0.25$ using the Cr\textsc{ii} 2056\,\AA\ and 2062\,\AA\ lines. We identified two Ly$\alpha$ absorption components which are potentially associated with the submm source Q0142$-$100b at $z=2.737$. We measured the H\textsc{i} column densities of these two systems to be log\,N(H\textsc{i}) $=15.0\pm0.25$ and $14.0\pm0.25$ (at $z=2.73579$ and $z=2.73701$, respectively).
\newline
\linebreak
\textbf{Q0207$-$003} -- 
There are at least two companion galaxies within a few arcseconds of Q0207$-$003 that have been detected in the near-IR, but these are not seen in our NOEMA observations.
\newline
\linebreak
\textbf{Q0449$-$1645} -- 
This source is only barely detected in our observations, and we found no companions in the surrounding field.
\newline
\linebreak
\textbf{Q1009$+$29} -- 
This source has a spectroscopically-detected companion about 6\,arcsec away. The companion galaxy has been identified at rest-frame optical, and has an unusual, diffuse morphology. This QSO also has a huge Ly\,$\alpha$ nebula surrounding it, extending all the way out to the companion. Due to the proximity of this companion to the host QSO, it is likely that the pair will become a merger.
\newline
\linebreak
\textbf{SBS1217$+$499} -- 
This source is well detected in our observations, and we found no companions in the surrounding field.
\newline
\linebreak
\textbf{HS1442$+$2931} -- 
The 850-$\mu$m flux density for this source was measured by Ross et al.~(in preparation) to be (2.6$\pm$0.9)\,mJy, which means that it has an $S_{850}/S_{3000}$ ratio of around 1.5. This is very low compared to the other sources in this sample. If we were to assume a thermal, modified blackbody spectral energy distribution with a temperature of 50\,K, redshift of 2.75 and emissivity index of 1.6, we would expect the flux density ratio \textbf{$S_{\rm 850\,\mu m}/S_{\rm 3\,mm}$} to be more like 2000; hence the 3-mm emission we have observed is certainly non-thermal. Being one of the faintest 850\,$\mu$m-emitters, one logical interpretation as to why we did not detect a CO(3--2) line would be that it is simply below our sensitivity. There are six spectroscopically confirmed galaxies within 30\,arcsec of the QSO and a handful of Ly\,$\alpha$-selected galaxies, evidence that this field is substantially over dense.
\newline
\linebreak
\textbf{HS1549$+$1919} -- This source has had the most extensive follow-up coverage of all the KBSS fields, and in particular about 200 galaxies have been discovered at the QSO redshift in the optical and IR within several arcminutes. One of these galaxies is seen in our NOEMA observations, while the other mm-detected galaxy does not have a counterpart; however, HS1549$+$1919 is the only QSO in this sample that clearly lies at the heart of a massive proto-cluster (Chapman et al.~in preparation), so it is not surprising that we have found yet another galaxy. The spectrum of this source contains two asymmetrical peaks, like those of HS1603$+$3820 and SBS1217$+$499, although here the secondary peak is much more prominent. Despite the extensive follow up observations, the secondary component has not been resolved. More details can be found in Chapman et al.~(in preparation).
\newline
\linebreak
\textbf{HS1603$+$3820} -- This field contains many Ly$\alpha$-selected galaxies and a 35-arcsec-scale Ly$\alpha$ nebula with AGN-like excitation and evidence for huge tidal tails, although none of these features are seen in our mm observations. 
\newline
\linebreak
\textbf{HS1700$+$64} -- 
Four submm sources from \citet{lacaille2018} at $z\approx2.31$ appear to have Ly$\alpha$ forest absorption near the redshifts of the systems along the HS1700$+$64 sightline. Based on redshifts alone, we cannot relate if this absorption is indeed associated with the submm sources. Nevertheless, the total H\textsc{i} column density measured over the full redshift range of the sources ($\Delta z=1360$\,km\,s$^{-1}$) is log\,N(H\textsc{i}) $=14.8\pm0.6$. No associated C\textsc{iv} is detected across the entire redshift span of these systems.
\newline
\linebreak
\textbf{Q2206$-$199} -- 
This field contains a companion galaxy with a strong CO(3--2) line that is notable for being quite asymmetrical. While we fit this line with a single Gaussian, the model is clearly not a great fit. We note that this companion is not seen in the optical or IR.
\newline
\linebreak
\textbf{Q2343$+$125} -- This source definitely stands out amongst this sample, being nearly 3 times brighter than the next brightest source. Interestingly, there is another QSO, FSzP1170, located only 5\,arcsec north of Q2343$+$125 at a redshift whose velocity difference corresponds to 670\,km\,s$^{-1}$, but that is undetected in our maps. Given the close proximity of FSzP1170 to Q2343$+$125 and the limited angular resolution of our NOEMA data (the angular separation makes the second QSO right on the border of our synthesized beam, and prominent sidelobes in the north-south direction nearly overlap with the position of FSzP1170), it could be that this secondary QSO is contributing to the high flux seen in Q2343$+$125. Future observations should follow up Q2343$+$125 in more detail in order to better understand this source. While the field does appear to contain a companion galaxy, we find that it does not contain a detectable CO(3--2) line. 

Towards the background quasar Q2343$+$125, we identified an intervening damped Ly$\alpha$ system at $z=2.3105$ with log\,N(H\textsc{i}) $=20.4\pm0.2$. Using the Si\textsc{ii} 1808\,\AA\ absorption, we estimate the metallicity of this system to be [Si/H] $=-0.59\pm0.25$. We note that this system is one of the most metal-rich systems identified to date \citep{Berg15}.

\end{document}